\documentclass[12pt]{article}
\usepackage{graphicx}
\usepackage[margin=1in]{geometry}
\usepackage{float}
\usepackage{graphicx}
\usepackage{multirow}
\usepackage[dvipsnames, usenames]{color}
\usepackage{tabularx}
\usepackage{algorithm, algpseudocode}
\usepackage{amsthm}
\usepackage{amssymb}
\usepackage{graphicx}
\usepackage{mathtools}
\usepackage{booktabs}
\usepackage{epstopdf}
\usepackage{array} 
\usepackage{booktabs}
\usepackage{natbib}
\usepackage{amsfonts,amsmath,amssymb}
\usepackage{enumerate}
\usepackage{bm}
\usepackage{mathrsfs}
\usepackage{mathtools}
\usepackage{booktabs}
\usepackage{array} 
\usepackage{subfig}
\usepackage{dsfont}
\usepackage{url}
\usepackage{lineno}

\newcommand{\calA}{\mathcal A}
\newcommand{\calB}{\mathcal B}

\newcommand{\calL}{\mathcal L}

\newcommand{\ba}{ {\boldsymbol a} }

\newcommand{\bK}{ {\boldsymbol K} }

\newcommand{\bu}{ {\boldsymbol u} }

\newcommand{\bv}{ {\boldsymbol v} }
\newcommand{\bV}{ {\boldsymbol V} }
\newcommand{\bw}{ {\boldsymbol w} }

\newcommand{\bx}{ {\boldsymbol x} }
\newcommand{\bX}{ {\boldsymbol X} }
\newcommand{\by}{ {\boldsymbol y} }

\newcommand{\bone}{ {\bf 1} }
\newcommand{\bzero}{ {\bf 0} }

\newcommand{\given}{\,|\,}

\newcommand{\balpha}{ {\boldsymbol \alpha} }

\newcommand{\bbeta}{ {\boldsymbol \beta} }

\newcommand{\bepsilon}{ {\boldsymbol \epsilon} }

\newcommand{\bet}{ {\boldsymbol \eta} }

\newcommand{\bOmega}{ {\boldsymbol \Omega} }

\newcommand{\btheta}{ {\boldsymbol \theta} }

\newcommand{\iid}{\overset{\mathrm{iid}}{\sim}}

\usepackage{setspace}
\newcommand{\subfigimg}[3][,]{%
  \setbox1=\hbox{\includegraphics[#1]{#3}}
  \leavevmode\rlap{\usebox1}
  \rlap{\hspace*{40pt}\raisebox{\dimexpr\ht1-2\baselineskip}{#2}}
  \phantom{\usebox1}
}

\newcommand{\beginsupplement}{%
        \setcounter{table}{0}
        \renewcommand{\thetable}{S\arabic{table}}%
        \setcounter{figure}{0}
        \renewcommand{\thefigure}{S\arabic{figure}}%
     }

\begin{document}   




\begin{center}
{\large Models to support forest inventory and small area estimation using sparsely sampled LiDAR: A case study involving G-LiHT LiDAR in Tanana, Alaska}
\vspace{0.5cm}

Andrew O. Finley$^{1,2}$, Hans-Erik Andersen$^{3}$, Chad Babcock$^{4}$, Bruce D. Cook$^{5}$,\\Douglas C. Morton$^{5}$, and Sudipto Banerjee$^{6}$
  
\vspace{5mm}
{\scriptsize
  $^1$Department of Forestry, Michigan State University, East Lansing, MI, USA.\\
  $^2$Department of Statistics and Probability, Michigan State University, East Lansing, MI, USA.\\
  $^3$USDA Forest Service,  Pacific Northwest Research Station, Seattle, WA, USA.\\
  $^4$Department of Forest Resources, University of Minnesota, St Paul, MN, USA\\
  $^5$National Aeronautics and Space Administration, Goddard Space Flight Center, Greenbelt, MD, USA.\\
  $^6$Department of Biostatistics, University of California, Los Angeles, CA, USA.
  
  \vspace{0.5cm}
  \textbf{Corresponding Author}: Andrew O. Finley, finleya@msu.edu
  }
\end{center}

\section*{Abstract}

A two-stage hierarchical Bayesian model is developed and implemented to estimate forest biomass density and total given sparsely sampled LiDAR and georeferenced forest inventory plot measurements. The model is motivated by the United States Department of Agriculture (USDA) Forest Service Forest Inventory and Analysis (FIA) objective to provide biomass estimates for the remote Tanana Inventory Unit (TIU) in interior Alaska. The proposed model yields stratum-level biomass estimates for arbitrarily sized areas. Model-based estimates are compared with the TIU FIA design-based post-stratified estimates. Model-based small area estimates (SAEs) for two experimental forests within the TIU are compared with each forest's design-based estimates generated using a dense network of independent inventory plots. Model parameter estimates and biomass predictions are informed using FIA plot measurements, LiDAR data that are spatially aligned with a subset of the FIA plots, and complete coverage remotely detected data used to define landuse/landcover stratum and percent forest canopy cover. Results support a model-based approach to estimating forest variables when inventory data are sparse or resources limit collection of enough data to achieve desired accuracy and precision using design-based methods. 



\section{Introduction}\label{sec:intro}

Large scale forest monitoring programs have traditionally used design-based inference that uses probability sampling and associated estimators to deliver forest variable estimates. For example, the United States Department of Agriculture (USDA) Forest Service Forest Inventory and Analysis (FIA) program conducts the US national forest inventory (NFI), collecting data describing the condition of forest ecosystems on a large network of permanent inventory plots distributed across all lands in the nation \citep{smith2002forest}. These data offer a unique and powerful resource for determining the extent, magnitude, and causes of long-term changes in forest health, timber resources, and forest landowner characteristics across large regions in the US \citep{wurtzebach2020supporting}. The FIA program uses design-based post-stratified estimators to improve precision of point and change estimates \citep{westfall2011post, bechtold2005enhanced}. Depending on the desired level of estimate precision, such approaches often require costly measurements over a relatively dense network of inventory plots; hence, from a cost efficiency standpoint, there is interest in methods that can deliver comparable inference using fewer inventory plots. At the same time, like other NFIs \citep{breidenbach2012small, kohl2006sampling}, FIA has experienced increased demand for estimates within smaller spatial, temporal, and biophysical extents than design-based inference can reasonably deliver (e.g., annual or stand-level estimates). Developing estimation methods that support inference on small areas---referred to as small area estimation (SAE) methods---using FIA data is an active area of research, with considerable progress made in the last several years \citep{hou2021updating, coulston2021enhancing, schroeder2014improving, lister2020use}. SAE methods are numerous and diverse, though most seek to improve inference on small areas by making use of statistical models and auxiliary information that is correlated with outcome variables \citep{rao2015small}.

The FIA and similar NFIs continue to explore the efficacy of model-assisted and model-based modes of inference to reduce cost and improve precision for both large and small area estimates. These approaches often leverage rich information content of satellite and airborne remote sensing to augment information gleaned from the inventory plot network. Model-assisted approaches employing complete coverage auxiliary (e.g., remote sensing) information to improve the precision of inventory estimates within the design-based inferential paradigm have been presented \citep{breidt2017model, mcconville2017model, strunk2019large, magnussen2018lidar, magnus2021}, while model-based techniques have been developed to provide estimates in cases where useful models are available to relate remote sensing metrics to inventory measurements \citep{Stahl2010, mcroberts2010probability, Finley2011, Saarela2016, babcock2018geostatistical, May2023}.

Model-based SAE methods offer a valuable alternative to the design-based post-stratified estimators implemented by FIA. Model-based SAE methods seek to borrow information from outside the area of interest (e.g., from neighboring areal-units or point-referenced observations) and auxiliary data (e.g., remote sensing data) to improve precision of estimated quantities. SAE methods can generally be classified into two groups: unit-level and area-level models. Unit-level models are constructed at the level of population units, where a population unit is defined as the minimal unit that can be sampled from a population. With respect to FIA's survey design, field plot centers represent population units. Unit-level models typically relate outcome variable measurements on sampled population units to auxiliary data that is available for all population units (e.g., complete coverage remote sensing data). Prediction for a small area is achieved by aggregating unit-level predictions within the given areal extent \citep{rao2015small}. In contrast, area-level models are constructed across areal units where relationships are built between area-specific outcome variable direct estimates (e.g., generated using design-based estimators) and auxiliary data \citep{rao2015small}. Hence, area-level models effectively ``adjust'' direct estimates given auxiliary information. 

The models developed and implemented in this paper are motivated by FIA's objective to improve biomass estimates in interior Alaska. FIA plot measurements in interior Alaska's Tanana Inventory Unit (TIU) were carried out over four years starting in 2014. Implementing FIA's standard sampling intensity is cost-prohibitive in this remote region, due to challenging logistics and high transportation costs---lack of roads requires that virtually every plot is accessed via helicopter. For this reason, FIA has implemented a modified sampling design in interior Alaska using a reduced sampling intensity for field plots (1 plot per 12\,140 ha), supplemented with high-resolution airborne imagery acquired by NASA Goddard's Lidar, Hyperspectral, and Thermal (G-LiHT) Airborne Imager \citep{cook2013nasa} in a sampling mode \citep{cahoon2022tanana}. Given the large geographic expanse of interior Alaska (approx. 46 million ha of forestland), a strip sampling mode is considered to be the most economical and spatially balanced acquisition strategy for the airborne data. The use of high-resolution airborne LiDAR strip sampling to support forest inventories has been the focus of several recent studies in Europe and North America where a variety of estimation approaches have been developed and evaluated, including model-assisted \citep{andersen2009estimating, gregoire2011model, strunk2019large}, model-based \citep{saarela2016hierarchical, staahl2011model}, and hybrid estimation \citep{stahl2016}. A comprehensive account of model-based inference for survey data and its richness over design-based estimators is offered by \cite{little2004jasa}, which includes specific pitfalls of the Horwitz-Thompson estimator. Recent explorations into Bayesian survey sampling from spatially correlated frameworks include \cite{Chan-Golston2020} and \cite{changolston2022jjsds} who show the benefits of modeling finite populations as realizations of a spatial process with the latter also accommodating modeling spatial associations in sampling indicators \citep[also see][]{Finley2011}.   

Here, we apply a model-based approach to estimate forest biomass within the TIU. The devised model yields stratum specific biomass estimates for arbitrarily sized areas. We generate biomass estimates for the entire TIU and two small areas. Model-based estimates are compared with the TIU FIA design-based post-stratified estimates. Model-based SAEs for two experimental forests within the TIU are compared with each forest's design-based estimates generated using a dense network of independent inventory plots (i.e., independent, meaning the data were not used to inform the model-based estimates). Model parameter estimates and biomass predictions are informed using FIA plot measurements, LiDAR data that is spatially aligned with a subset of the FIA plots, and complete coverage remotely sensed data used to define four landuse/landcover stratum and percent forest canopy cover. Model data input and subsequent biomass predictions are point-referenced (i.e., indexed by spatial locations) hence, from a SAE standpoint, we pursue a unit-level approach.

The remainder of the paper is as follows. Section~\ref{sec:data} provides a description of the data and exploratory analysis. The modeling framework and assessment criteria are described in Section~\ref{sec:models}. Design-based estimators used to generate estimates to compare with model-based estimates are provided in Section~\ref{sec:design}. A description of the TIU and small area analyses are provided in Section~\ref{sec:analyses}. Model selection results and biomass estimates for the TIU and small areas are presented in Section~\ref{sec:results}. Results and description of possible next steps are discussed in Section~\ref{sec:discussion}. 

\section{Data}\label{sec:data}

\begin{figure}[ht!]
\centering
\subfloat[]{\includegraphics[width=7cm,trim={0cm 2.5cm 0cm 2cm},clip]{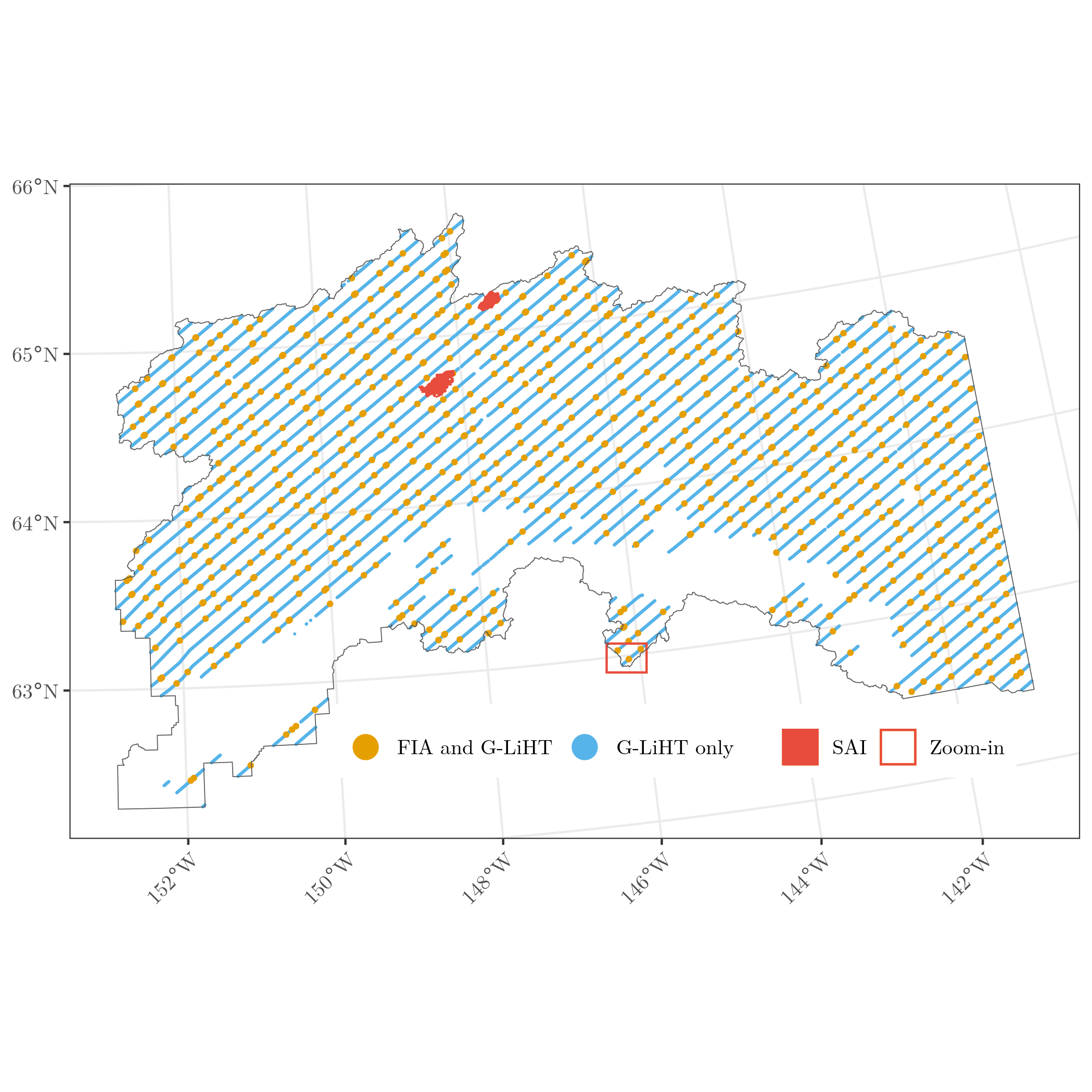}\label{fig:tiu-data}}
\subfloat[]{\includegraphics[width=7cm,trim={0cm 2.5cm 0cm 2cm},clip]{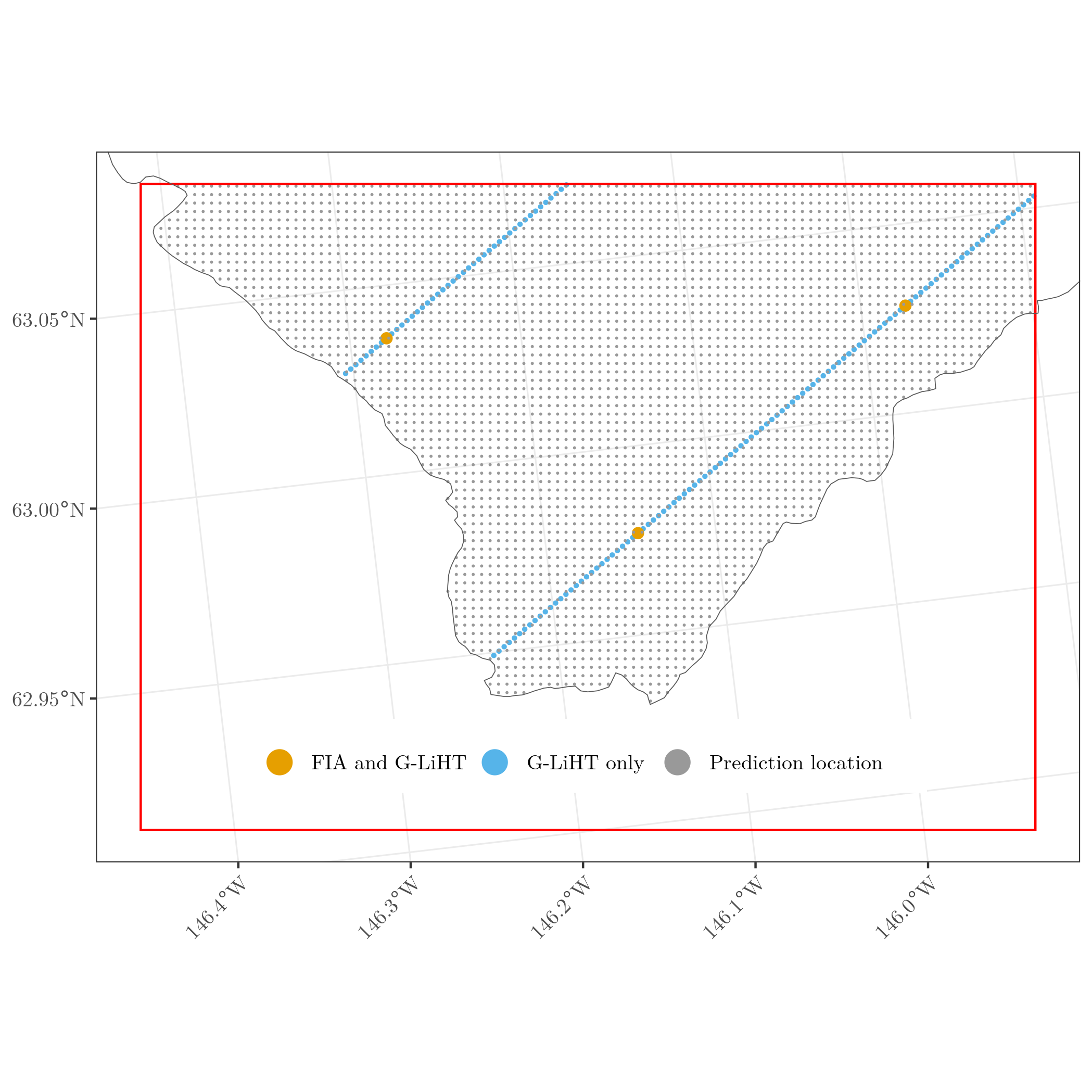}\label{fig:tiu-data-zoom}}\\
\subfloat[]{\includegraphics[width=7cm,trim={0cm 2.5cm 0cm 2cm}, clip]{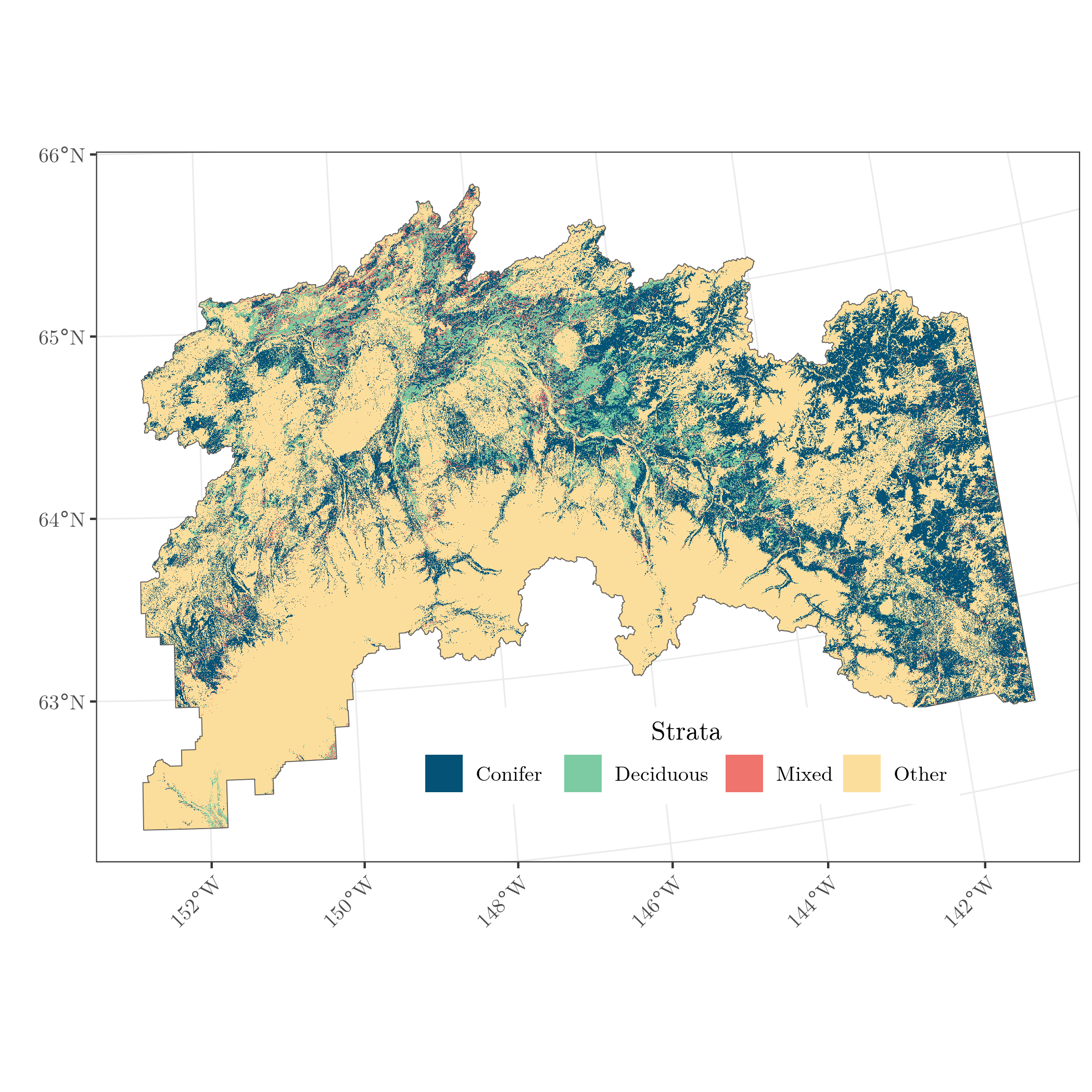}\label{fig:tiu-strata}}
\subfloat[]{\includegraphics[width=7cm,trim={0cm 2.5cm 0cm 2cm},clip]{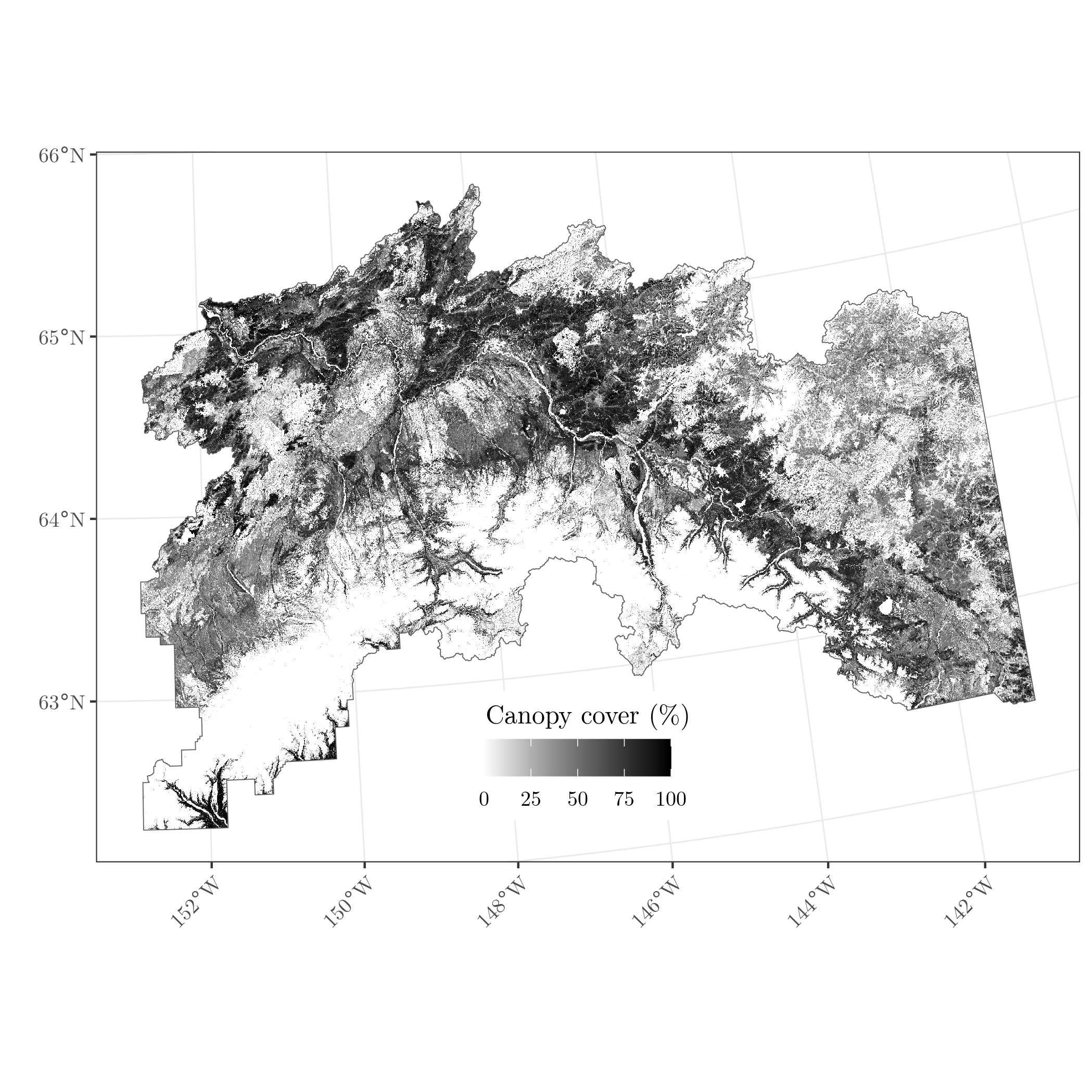}\label{fig:tiu-hansen}}
\caption{\protect\subref{fig:tiu-data} Tanana Inventory Unit with data collection locations (FIA locations are perturbed) and two small areas of interest (SAI). \protect\subref{fig:tiu-data-zoom} Zoom-in that shows all three location types (FIA locations are perturbed). The dense grid of prediction locations exist across the entire TIU. \protect\subref{fig:tiu-strata} Strata used in subsequent modeling. \protect\subref{fig:tiu-hansen} Percent forest cover.}\label{fig:tiu}
\end{figure}

We describe TIU data used to inform the proposed biomass models presented in Section~\ref{sec:models}. The 13.533 million (ha) TIU is shown in Figure~\ref{fig:tiu-data} along with data locations and extents.

\subsection{FIA plots}\label{sec:fia}

The standard FIA plot comprises four 1/60-th (ha) fixed-area, circular subplots spaced 36.6 (m) apart \citep[see][for details on FIA plot design and measurements]{cahoon2022tanana}. Subplot coordinates were obtained using a GLONASS-enabled Trimble GeoXH mapping-grade GNSS receivers that provide $<$ 2 (m) geolocation error \citep{mcgaughey2017effect, andersen2009accuracy, andersen2022using}. Following FIA procedures, individual tree dry biomass was estimated, summarized to the plot-level, and expressed on a per ha basis \citep[see FIA DRYBIOT variable in][]{FIADB}. This plot-level summarized and expanded DRYBIOT variable is $y$ (Mg/ha) in subsequent model development.

As noted previously, TIU FIA plot measurements were collected over four years (2014, 2016-2018). Over this period, 1\,091 FIA plots were sampled. The subset of 880 plots that spatially align with G-LiHT data (described in the Section~\ref{sec:gliht}) are depicted in Figure~\ref{fig:tiu-data}. This subset of plots is used in subsequent model development, whereas all sampled plots are used for subsequent design-based estimates. To protect the plot integrity and ownership information, FIA plot location is proprietary. While actual locations were used for modeling, figures presented here depict spatially perturbed (i.e., fuzzed) locations.

\subsection{G-LiHT LiDAR}\label{sec:gliht}

To augment sparsely sampled FIA plots, linear swaths of high-resolution airborne remote sensing measurements, placed approximately 9.2 (km) apart and spatially aligned with most of the FIA plots, were acquired in 2014 using G-LiHT mounted on a fixed-wing aircraft platform \citep{cook2013nasa}. G-LiHT data specifications are provided in Table~\ref{tab:gliht}.

\begin{table}[H]
\centering
\begin{tabular}{ll|ll}
Specification&&Specification&\\
\midrule
Instrument & Riegl VQ-480 &Footprint size & 10 cm\\
Laser wavelength & 1550 nm &Half-scanning angle & 15 degrees\\
Flying height & 3 35 m (AGL) &Average pulse density & 3 pulses/m$^2$ \\
Beam divergence & 0.3 mrad &Swath width & 400 m\\
\bottomrule
\end{tabular}\caption{G-LiHT LiDAR specifications for TIU.}\label{tab:gliht}
\end{table}

A 1-by-1 (m) canopy height model (CHM) was computed as the difference between G-LiHT derived digital terrain model (DTM) and a canopy surface model (CSM). The average canopy height (m) was computed from the CHM over the 880 spatially coinciding FIA plot footprints and a series of 61\,029 ``LiDAR plots'' each with area equal to the FIA plot and spaced 200 (m) apart along the linear flight lights. The average CHM height variable is $x_{CH}$ in subsequent analyses.

G-LiHT LiDAR plot locations are shown in Figures~\ref{fig:tiu-data} and \ref{fig:tiu-data-zoom}. We could have tessellated the G-LiHT swath into 10s of millions of FIA plot sized LiDAR plots; however, results from \cite{Finley2019}, \cite{shirotaconjugate}, and \cite{Michele2021arxiv} show there is strong spatial dependence among G-LiHT canopy height metrics in TIU and hence little additional information gain at the expense of massive model fitting computational cost.

\subsection{Stratification and forest canopy cover}\label{sec:complete_cover}

A complete coverage (i.e., covering the entire TIU) 30-by-30 (m) resolution stratification variable was formed using forest and non-forest National Land Cover Database (NLCD) \citep{homer2015completion} with stratum ``Deciduous'' (Class 41), ``Conifer'' (Class 42), ``Mixed'' (Class 43), and ``Other'' (all non-forest classes) (Figure~\ref{fig:tiu-strata}). Using a model-assisted approach to analyze the TIU data described above, \cite{Andersen2023} found this NLCD stratifying variable explained a substantial portion of variability in biomass. Hence, we too use this stratification variable to facilitate comparison with \cite{Andersen2023}.

In addition to the NLCD stratification variable, a complete coverage 30-by-30 (m) resolution percent tree cover variable was formed using \cite{Hansen2013} fractional tree cover, updated to account for forest cover loss prior to 2014 (Figure~\ref{fig:tiu-hansen}). This tree cover variable is $v_{TC}$ in subsequent analyses.

\subsection{Exploratory data analysis}\label{sec:eda}

In Section~\ref{sec:models}, we define a hierarchical Bayesian regression model used to predict biomass ($y$) using NLCD strata and G-LiHT $x_{CH}$. As with all modeling efforts, we begin with exploratory data analysis (EDA) to build intuition about the data and relationships of interest.

\begin{figure}[ht!]
\centering
\subfloat[]{\includegraphics[width=7cm,trim={0cm 0cm 0cm 0cm},clip]{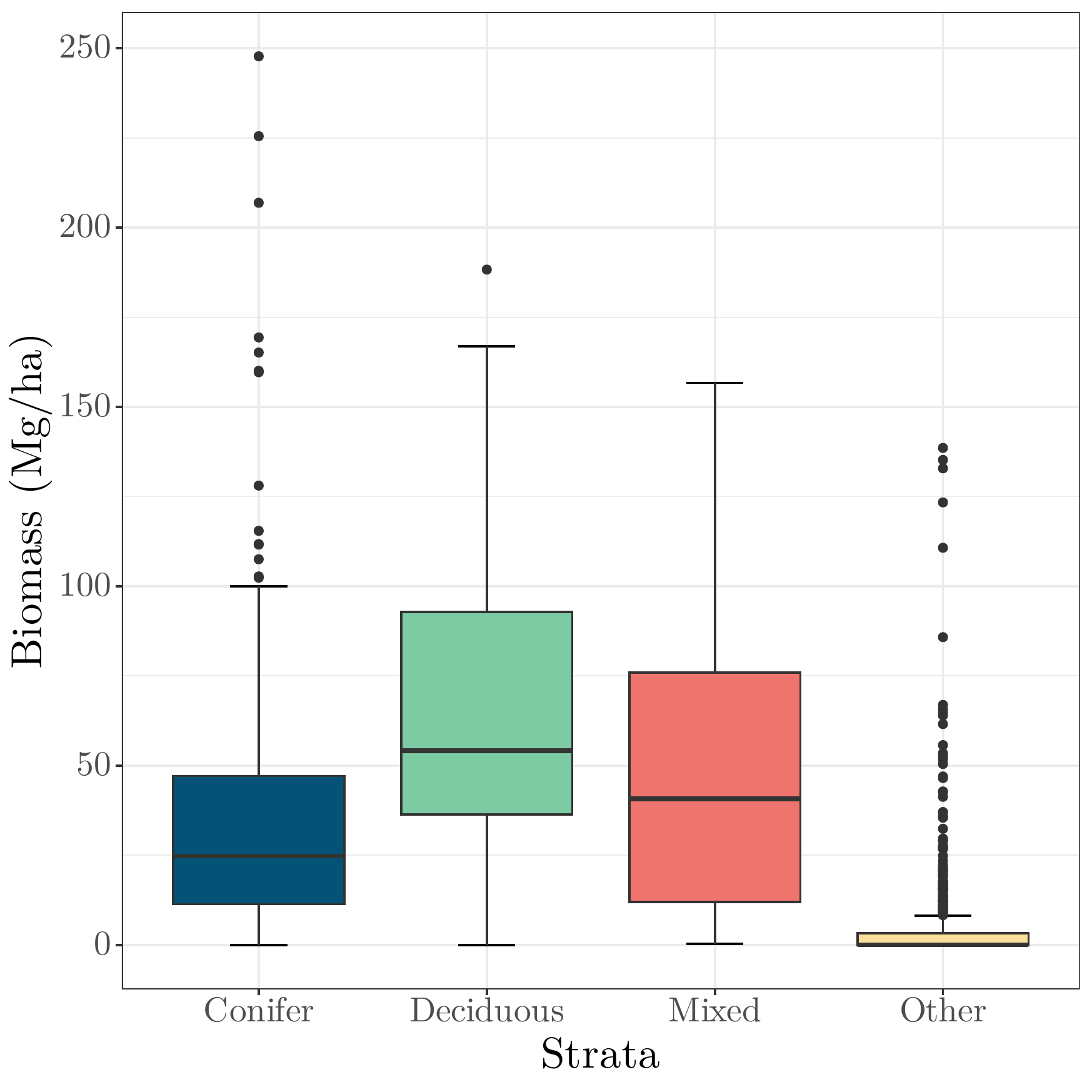}\label{fig:boxplots-bio}}
\subfloat[]{\includegraphics[width=7cm,trim={0cm 0cm 0cm 0cm},clip]{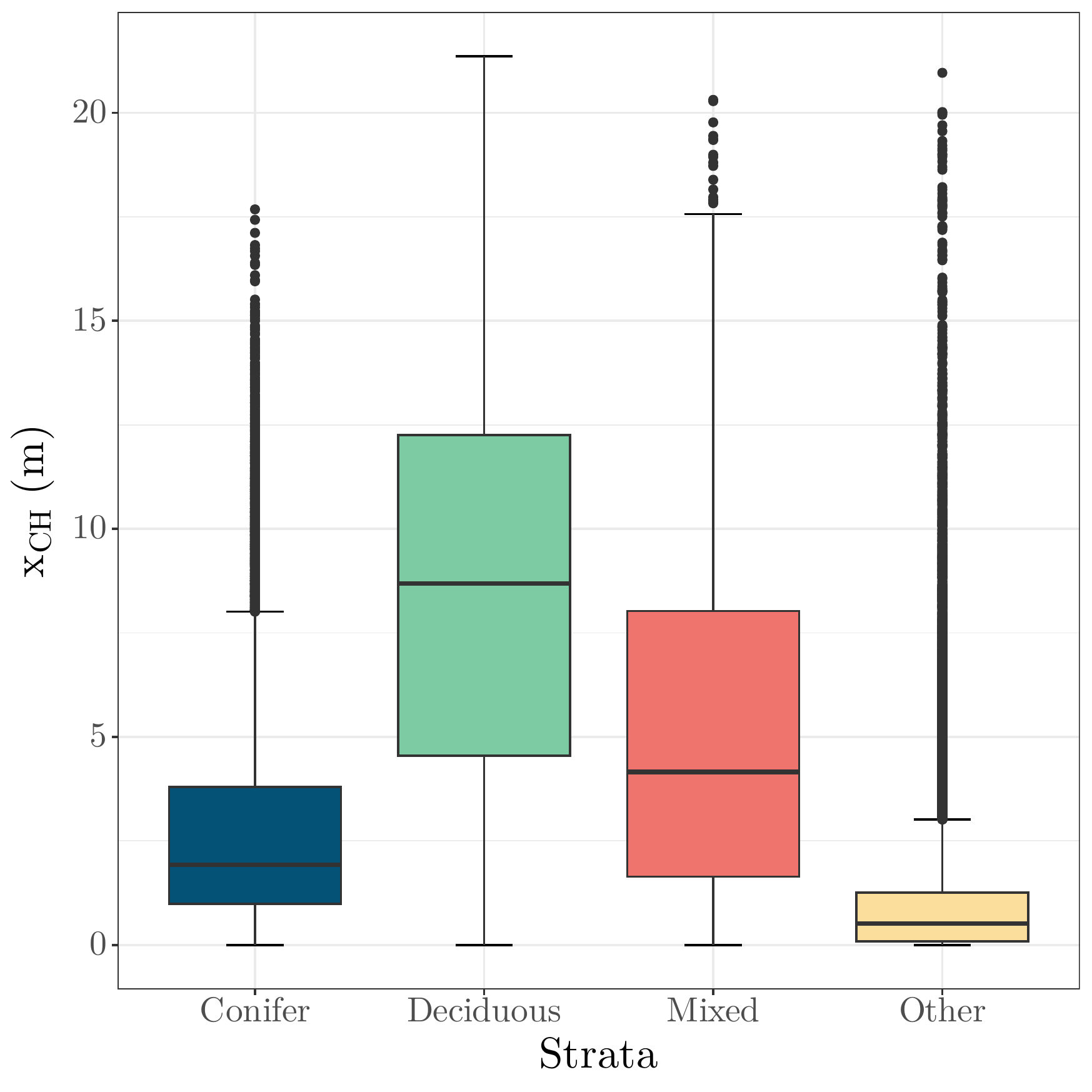}\label{fig:boxplots-rs}}
\caption{\protect\subref{fig:boxplots-bio} distribution of biomass and \protect\subref{fig:boxplots-rs} G-LiHT mean canopy height $x_{CH}$ by NLCD strata. Horizontal boxplot lines indicate the distribution's quantiles and points identify extreme measurements.}\label{fig:boxplots}
\end{figure}

Figure~\ref{fig:boxplots-bio} shows the distribution of FIA plot biomass density measurements is different for each stratum. Conifer forests have lower median biomass than Deciduous or Mixed forest types, but also several plots with very high biomass (i.e., right skewed distribution). Among the forested strata, Deciduous has the largest median biomass. As the stratum name suggests, biomass distribution within Mixed appears to be a mix of Conifer and Deciduous distributions. As described in Section~\ref{sec:complete_cover}, the Other stratum is a catch-all for non-forest classes. Across the TIU, non-forest landcover is dominated by barren, herbaceous, and wetlands, hence the forest biomass distribution in Other is concentrated just above zero. The substantial right skew in Other is due to forested FIA plots that fall within non-forest NLCD classes.    
Figure~\ref{fig:boxplots-rs} shows distribution of mean forest canopy height, summarized using $x_{CH}$, is also different for each stratum. As observed in numerous studies, forest biomass and forest canopy structure metrics, like $x_{CH}$, have a positive correlation, therefore it is not surprising the patterns seen in Figure~\ref{fig:boxplots-bio} are reflected in Figure~\ref{fig:boxplots-rs}.

\begin{figure}[ht!]
\centering
\includegraphics[width=8cm]{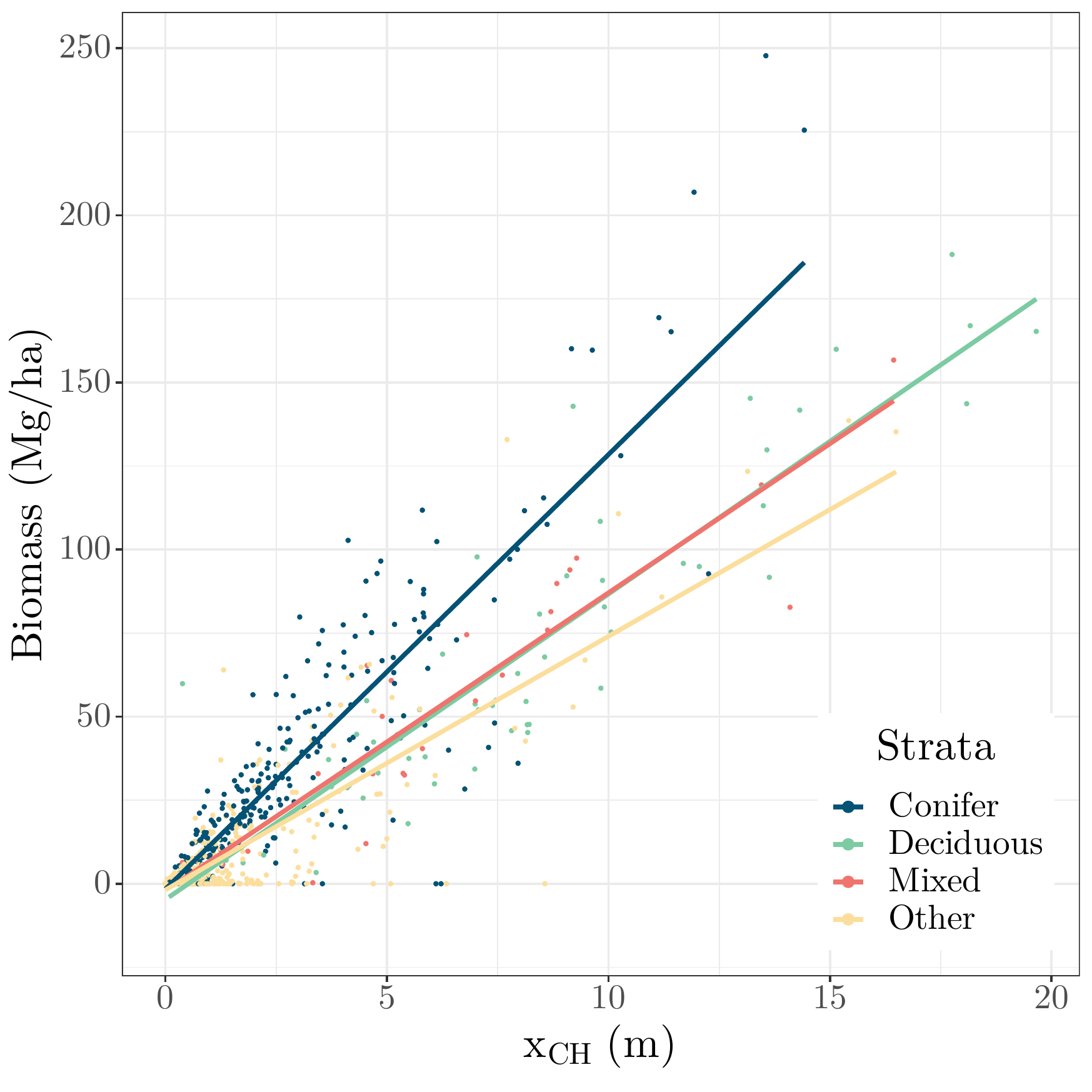}
\caption{Biomass and G-LiHT derived mean canopy height $x_{CH}$ observed at FIA plot locations. Model (\ref{eq:lm}) regression lines correspond to parameter estimates given in Table~\ref{tab:edaEsts}. }\label{fig:edaReg}
\end{figure}

Next we explore the relationship between biomass and mean forest canopy height by stratum using the simple linear regression
\begin{equation}\label{eq:lm}
    y = \beta_0 + \beta_{CH}x_{CH} + \epsilon,
\end{equation}
where $\beta_0$ and $\beta_{CH}$ are the intercept and slope coefficient, respectively, and $\epsilon \iid N(0, \tau^2)$. This model was fit to stratum specific data. These data and resulting regression lines are shown in Figure~\ref{fig:edaReg}. Sample size, $n$, and parameter estimates are given in  Table~\ref{tab:edaEsts}. Results suggest a strong linear relationship between biomass and $x_{CH}$ within each stratum. Among the forested strata, the regression slope coefficient for Conifer is larger than Deciduous and Mixed. The slope coefficient for Other is substantially smaller than forested strata coefficients. These differences suggest subsequent model development could benefit from stratum-varying coefficients. 

The last row in Table~\ref{tab:edaEsts} shows parameter estimates for the pooled model (i.e., not broken out by stratum). Given stratum specific differences in $\tau^2$ estimates and magnitude of the pooled model's $\tau^2$ estimate compared with that of the Other's model (i.e., $\sim$251 vs. $\sim$85), we might expect subsequent model development could benefit from stratum specific residual variance parameters. 

\begin{table}[ht!]
\begin{center}
{\small
\begin{tabular}{lcccc}
 & &$\beta_0$  & $\beta_{CH}$ & $\tau^2$\\
\cmidrule(lr){3-3}
\cmidrule(lr){4-4}
\cmidrule(lr){5-5}
Stratum & $n$ & $_{\text{(L. 95\%)}}\;\;\text{Mean}\;\;_{\text{(U. 95\%)}} $& $_{\text{(L. 95\%)}}\;\;\text{Mean}\;\;_{\text{(U. 95\%)}} $& $_{\text{(L. 95\%)}}\;\;\text{Mean}\;\;_{\text{(U. 95\%)}} $ \\
\midrule
Conifer&265&$_{(\text{-}5.206)}\;\;\text{-}1.616\;\;_{(1.986)}$&$_{(12.059)}\;\;13.008\;\;_{(13.923)}$&$_{(288.871)}\;\;341.878\;\;_{(403.191)}$\\
 Deciduous&29&$_{(\text{-}15.817)}\;\;\text{-}4.501\;\;_{(6.070)}$&$_{(7.973)}\;\;9.141\;\;_{(10.333)}$&$_{(283.412)}\;\;413.825\;\;_{(609.904)}$\\
 Mixed&56&$_{(\text{-}12.645)}\;\;\text{-}2.328\;\;_{(7.459)}$&$_{(7.619)}\;\;8.952\;\;_{(10.435)}$&$_{(141.354)}\;\;226.322\;\;_{(396.705)}$\\
 Other&530&$_{(\text{-}2.765)}\;\;\text{-}1.906\;\;_{(\text{-}0.948)}$&$_{(7.161)}\;\;7.596\;\;_{(8.022)}$&$_{(76.062)}\;\;85.039\;\;_{(96.779)}$\\
 \midrule
Pooled&880&$_{(\text{-}2.464)}\;\;\text{-}1.166\;\;_{(0.132)}$&$_{(9.518)}\;\;9.866\;\;_{(10.212)}$&$_{(227.651)}\;\;251.259\;\;_{(275.621)}$\\
 \bottomrule
\end{tabular}
}
\caption{Sample size $n$ and parameter estimates for stratum specific model (\ref{eq:lm}) fit using biomass and G-LiHT derived mean canopy height $x_{CH}$ observed at FIA plot locations. Data and regression lines are shown in Figure~\ref{fig:edaReg}.}\label{tab:edaEsts}
\end{center}
\end{table}

Regression model diagnostic analyses identified a few potentially high leverage observations and moderate residual heteroskedasticity. Additionally, semivariogram analysis suggested negligible spatial dependence among model residuals (i.e., spatial dependence among biomass measurements was, for the most part, captured using $x_{CH}$). Previous TIU modeling efforts conducted using FIA subplots, which allow for better estimate of the semivariogram nugget due to closer spatial proximity, and denser plot networks (see, e.g., \citealp{babcock2018geostatistical, TaylorRodriguez2019, shirotaconjugate}), showed that even after accounting for LiDAR derived forest canopy variables, residual spatial dependence in biomass was present, and model fit and predictive performance was improved via addition of spatial random effects. 

\section{Models}\label{sec:models}

We propose a model to predict biomass density at any location(s) within the TIU. The model accommodates stratum-varying regression intercept, slope, and variance parameters, as well as residual spatial dependence. Biomass prediction for an area (e.g., entire TIU or small area within the TIU) is approximated by summarizing an appropriately dense grid of unit-level predictions within the area. For example, biomass density and total estimates for the TIU will be based on posterior predictive distributions estimated at each location within a 250-by-250 (m) grid that extends across the TIU---a portion of which is illustrated in Figure \ref{fig:tiu-data-zoom}. In our setting, a key issue with this approach is that biomass is to be conditioned on G-LiHT derived canopy structure metrics, e.g., $x_{CH}$, and these metrics are observed only along the flight lines. This misalignment is remedied using a two-stage approach. The first stage is a biomass process model that is conditioned on canopy structure metrics, and the second stage  comprises a separate process model for each canopy structure metric. For the first stage model, we assume all biomass measurements have spatially coinciding canopy structure measurements. When cast within a hierarchical Bayesian framework, this two-stage approach allows uncertainty in canopy structure metric predictions to be propagated through to biomass predictions.

Let us consider a first stage biomass model at generic location $\ell$ within the TIU. We model biomass $y(\ell)$ using a set of $p$ canopy structure variables $\bx(\ell)$ measured using G-LiHT. Based on the exploratory analysis, we allow the relationship between these canopy structure variables and biomass to vary by stratum via $p$ stratum-varying coefficients $\tilde{\bbeta}_j$ for $j$ in $1, 2, \ldots, q$, where $q$ is the number of strata. Additionally, an underlying latent spatial process $w(\ell)$ accounts for local changes in biomass attributed to unobserved, but smoothly varying, environmental influences. Note that a spatial location $\ell$ resides in one specific stratum, which implies that the stratum index $j$ is completely determined by $\ell$. The biomass $y(\ell)$ for a location $\ell$ situated inside the $j$-th stratum is modeled as
\begin{equation}
y(\ell) = \beta_0 + \tilde{\beta}_{0}(\ell) + \bx(\ell)^\top(\bbeta + \tilde{\bbeta}(\ell)) + w(\ell) + \epsilon(\ell), 
\label{eq:mody}
\end{equation}
where $\beta_0$ is the intercept, $\tilde{\beta}_{0}(\ell) = \tilde{\beta}_{0,j}$ for all $\ell$ in stratum $j$, $\bx(\ell)$ is the $p\times 1$ vector of canopy structure variables with associated global regression coefficients $\bbeta$ and stratum-specific coefficients $\tilde{\bbeta}(\ell) = \tilde{\bbeta}_j$ for all $\ell$ in stratum $j$, $w(\ell)$ is a spatial random effect that adds local adjustment with spatial dependence, and the residual is modeled as $\epsilon(\ell) \overset{ind.}{\sim} N(0, \tau^2(\ell))$ with $\tau^2(\ell) = \tau^2_j$ for all $\ell$ in stratum $j$. Stratum effects are modeled as $\tilde{\beta}_{0,j} \iid N(0, \sigma^2_0)$ and $\tilde{\bbeta}_{j} = (\tilde{\beta}_{1,j}, \tilde{\beta}_{2,j}, \ldots, \tilde{\beta}_{p,j})^\top$ with $\tilde{\beta}_{k,j} \iid N(0, \sigma^2_k)$ for $k$ in $1, 2, \ldots, p$. The spatial random effect is modeled as a Nearest Neighbor Gaussian Process (NNGP) $w(\ell) \sim NNGP(0, \sigma^2_w \rho(\cdot, \cdot; \phi_w))$, where $\sigma^2_w$ is the variance, $\rho$ is a spatial correlation function defined for pairs of locations within the domain, and $\phi_w$ captures correlation between locations based on spatial separation \citep[see, e.g.,][]{Datta2016, Banerjee2017}. Briefly, the NNGP specification implies the spatial random effect vector over $n$ locations $\bw = (w(\ell_1), w(\ell_2), \ldots, w(\ell_n))^\top$ has probability distribution $N(\bzero, \bK_w)$, where $\bK_w$ is an $n\times n$ NNGP covariance matrix derived from the covariance function $K_w(\ell, \ell'; \btheta_w) = \sigma_w^2\rho(\ell, \ell'; \phi_w) = \sigma_w^2\exp(-\phi_w||\ell - \ell'||)$ where $\exp(\cdot)$ is an exponential spatial correlation function, $\btheta_w = \{\sigma_w^2, \phi_w\}$, and $||\ell - \ell'||$ is the Euclidean distance between possibly different locations $\ell$ and $\ell'$. 

To allow prediction of $y(\ell)$ at locations where G-LiHT is not observed, each canopy structure variable is modeled similar to biomass and includes both stratum-varying regression coefficients and a continuous spatial process. Specifically, the second stage model for the $k$-th canopy structure variable at location $\ell$ situated inside the $j$-th stratum is
\begin{equation}
x_k(\ell) = \alpha_{k,0} + \tilde{\alpha}_{k,0}(\ell) + \bv_{k}(\ell)^\top(\balpha_k + \tilde{\balpha}_{k}(\ell)) + u_k(\ell) + \eta_{k}(\ell), 
\label{eq:modx}
\end{equation}
where $\alpha_{k,0}$ is the intercept, $\tilde{\alpha}_{k,0}(\ell) = \tilde{\alpha}_{k,0,j}$ for all $\ell$ in stratum $j$, $\bv_{k}(\ell)$ is a $r\times 1$ vector of predictors with associated global regression coefficients $\balpha_k$ and stratum-specific effects $\tilde{\balpha}_k(\ell) = \tilde{\balpha}_{k,j}$ for all $\ell$ in stratum $j$, $u_k(\ell)$ is a spatial random effect, and $\eta_{k}(\ell) \overset{ind.}{\sim} N(0, \gamma^2_{k}(\ell))$ with $\gamma^2_{k}(\ell) = \gamma^2_{k,j}$ for all $\ell$ in stratum $j$. Stratum effects are modeled as $\tilde{\alpha}_{k,0,j}\iid N(0, \nu^2_{k,0}$) and $\tilde{\balpha}_{k,j} = (\tilde{\alpha}_{k,1,j}, \tilde{\alpha}_{k,2,j}, \ldots, \tilde{\alpha}_{k,r,j})^\top$ with $\tilde{\alpha}_{k,i,j} \iid N(0, \nu^2_{k,i})$ for $i$ in $1, 2, \ldots, r$. The spatial random effect is modeled as $u_k(\ell)\sim NNGP(0, \nu^2_{k,u} \rho(\cdot, \cdot; \phi_{k,u}))$,  where $\nu^2_{k,u}$ is the variance, $\rho$ is again taken as an exponential spatial correlation function, and $\phi_{k,u}$ is the spatial decay parameter. We again collect the spatial process parameters into a vector $\btheta_{k,u} = \{\nu^2_{k,u}, \phi_{k,u}\}$.

For the set of $n$ locations $\calL = (\ell_1, \ell_2, \ldots, \ell_n)$ where both biomass and G-LiHT canopy structure data are observed, we define the $n\times 1$ vector $\by = (y(\ell_1), y(\ell_2), \ldots, y(\ell_n))^\top$, $n\times 1$ vector of ones $\bone$, $n\times q$ matrix $\mathds{1}$ with the $(i,j)$-th element equaling 1 if $\ell_i$ falls within the $j$-th stratum and zero otherwise, $n\times p$ matrix $\bX$ with row $i$ equal to $\bx(\ell_i)^\top$, $n\times pq$ matrix $\tilde{\bX}$ with row $i$ and columns $(j-1)p+1$ through $(j-1)p+p$ equal to $\bx(\ell_i)^\top$ when location $\ell_i$ falls within the $j$-th stratum and zero otherwise, $\bw = (w(\ell_1), w(\ell_2), \ldots, w(\ell_n))^\top$, and $\bepsilon = (\epsilon(\ell_1), \epsilon(\ell_2), \ldots, \epsilon(\ell_n))^\top$. We then write model (\ref{eq:mody}) as 
\begin{equation}
    \by = \beta_0\bone + \mathds{1}\tilde{\bbeta}_0 + \bX\bbeta + \tilde{\bX}\tilde{\bbeta} + \bw + \bepsilon,
\end{equation}
where $\beta_0$ and $\bbeta$ are as defined earlier, $\tilde{\bbeta}_0 = (\tilde{\beta}_{0,1}, \tilde{\beta}_{0,2}, \ldots, \tilde{\beta}_{0,q})^\top$, and $\tilde{\bbeta} = (\tilde{\bbeta}_1^\top, \tilde{\bbeta}_2^\top, \ldots, \tilde{\bbeta}_q^\top)^\top$. Let $\bOmega_y = \{\beta_0, \tilde{\bbeta}_0, \bbeta, \tilde{\bbeta}, \{\sigma^2_i\}_{i=0}^p, \bw, \btheta_w, \{\tau_j^2\}_{j=1}^q\}$ denote the collection of parameters in the above model including those specifying the prior distributions and the spatial process as described below (\ref{eq:mody}) to complete the hierarchical model.

Similarly, for the $n_s$ locations $\calL_s = (\ell_1, \ell_2, \ldots, \ell_{n_s})$ where G-LiHT is observed, we define the $n_s \times 1$ vector for the $k$-th canopy structure variable $\bx_k = (x_k(\ell_1), x_k(\ell_2), \ldots, x_k(\ell_{n_s}))^\top$, $n_s\times 1$ vector of ones $\bone_k$, $n_s\times q$ matrix $\mathds{1}_k$ with row $i$ and column $j$ equal 1 when $\ell_i$ falls within the $j$-th stratum and zero otherwise, $n_s \times r$ matrix $\bV$ with row $i$ equal to $\bv(\ell_i)^\top$, $n_s\times rq$ matrix $\tilde{\bV}$ with row $i$ and columns $(j-1)r+1$ through $(j-1)r+r$ equal to $\bv(\ell_i)^\top$ when location $\ell_i$ falls within the $j$-th stratum and zero otherwise, $\bu_k = (u_k(\ell_1), u_k(\ell_2), \ldots, u_k(\ell_{n_s}))^\top$, and $\bet_k = (\eta_k(\ell_1), \eta_k(\ell_2), \ldots, \eta_k(\ell_{n_s}))^\top$. We then write model (\ref{eq:modx}) as 
\begin{equation}
    \bx_k = \alpha_{k,0}\bone_k + \mathds{1}_k\tilde{\balpha}_{k,0} + \bV\balpha_k + \tilde{\bV}\tilde{\balpha}_k + \bu_k + \bet_k.
\end{equation}
where 
$\tilde{\balpha}_{k,0} = (\tilde{\alpha}_{k,0,1}, \tilde{\alpha}_{k,0,2}, \ldots, \tilde{\alpha}_{k,0,q})^\top$, and $\tilde{\balpha}_k = (\tilde{\balpha}_{k,1}^\top, \tilde{\balpha}_{k,2}^\top, \ldots, \tilde{\balpha}_{k,q}^\top)^\top$. The spatial random effect vector $\bu_k \sim NNGP(\bzero, \bK_{k,u})$, where $\bK_{k,u}$ is an $n_s\times n_s$ NNGP covariance matrix defined analogously to $\bK_w$. Model parameters are collected in the vector $\bOmega_{x_k} = (\alpha_{k,0}, \tilde{\balpha}_{k,0}, \balpha_k, \tilde{\balpha}_k, \{\nu^2_{k,i}\}_{i=0}^r, \bu_k, \btheta_{k,u}, \{\gamma^2_{k,j}\}_{j=1}^q)$, which includes parameters specifying prior distributions in spatial process as described below (\ref{eq:modx}).

To complete the Bayesian specification of these models, a prior distribution is assigned to each parameter. For model (\ref{eq:mody}) and (\ref{eq:modx}) we assign: flat priors to $\beta_0$, $\alpha_{k,0}$ and elements in $\bbeta$ and $\balpha_k$; implicit Normal distributions for elements in $\tilde{\bbeta}_0$, $\tilde{\bbeta}$, $\tilde{\balpha}_{k,0}$, and $\tilde{\balpha}_k$; NNGP distribution for $\bw$ and $\bu_k$; weakly informative inverse-Gamma distributions with hyperparameter shape equal to 2 and scale value guided by EDA results for all variance parameters; Uniform distributions with support between 1 to 500 (km) for decay parameters $\phi_w$ and $\phi_u$.  

Following the Bayesian mode of inference, we employ a Markov chain Monte Carlo (MCMC) algorithm to generate samples from model (\ref{eq:mody}) parameters' joint posterior distribution,
\begin{equation}
[\bOmega_y \given \by, \mathds{1}, \bX] \propto [\bOmega_y] \times [\by\given \bOmega_y, \mathds{1}, \bX], 
    \label{postMody}
\end{equation}
where $[\bOmega_y \given \by, \mathds{1}, \bX]$ represents the joint posterior distribution of $\bOmega_y$ conditioned on the data, $[\bOmega_y]$ is the joint prior distribution, and $[\by\given \bOmega_y, \mathds{1}, \bX]$ represents the likelihood. Parameter inference is obtained by sampling from (\ref{postMody}) using MCMC. After diagnosing convergence, we collect $L$ samples from (\ref{postMody}), which are denoted as $(\bOmega_y^{(1)}, \bOmega_y^{(2)}, \ldots, \bOmega_y^{(L)})$. Similarly, we obtain estimates for (\ref{eq:modx}) by sampling from the joint posterior distribution, 
\begin{equation}
[\bOmega_{x_k} \given \bx_k, \mathds{1}_k, \bV] \propto [\bOmega_{x_k}]\times [\bx_k\given \bOmega_{x_k}, \mathds{1}_k, \bV], 
    \label{postModx}
\end{equation}
with $L$ post-convergence MCMC samples collected as $(\bOmega_{x_k}^{(1)}, \bOmega_{x_k}^{(2)}, \ldots, \bOmega_{x_k}^{(L)})$.

\subsection{Posterior predictive inference and areal estimates}\label{sec:ppd}
As described in Section~\ref{sec:intro}, our inferential objective is to estimate biomass density (Mg/ha) and total (Mg) for any user-defined area within the TIU. This inference uses model (\ref{eq:mody}) to estimate the biomass density posterior predictive distribution at each of $n^\ast$ prediction locations, $\calL_0 = (\ell_{0,1}, \ell_{0,2}, \ldots, \ell_{0,n^\ast})$, laid in a dense grid over the area of interest. These predictions are then used to estimate the desired area summaries of biomass density and total. However, because G-LiHT canopy structure variables are not observed at prediction locations, we must condition biomass predictions from model (\ref{eq:mody}) on predictions of canopy structure variables from model (\ref{eq:modx}). Ideally, this is done in a way that propagates the uncertainty in canopy structure variable predictions through to biomass predictions.

For inference at unobserved locations, we extend model (\ref{eq:modx}) from observed locations to any arbitrary location $\ell_{0}$. If $\bx^\ast_{k} = (x^\ast_k(\ell_{0,1}), x^\ast_k(\ell_{0,2}), \ldots, x^\ast_k(\ell_{0,n^\ast}))^\top$ is the vector of unknown measurements and $\bu_k^{\ast}$ is the analogously defined vector of spatial random effects corresponding to the $k$-th canopy structure variable, the NNGP extends model (\ref{eq:modx}) such that if $\ell_{0,m}$ falls within the $j$-th stratum then
\begin{equation}\label{eq:modx_pred}
    x^{\ast}_k(\ell_{0,m}) = \alpha_{k,0} + \tilde{\alpha}_{k,0,j} + \bv_{k}(\ell_{0,m})^\top(\balpha_k + \tilde{\balpha}_{k,j}) + u_k(\ell_{0,m}) + \eta_k(\ell_{0,m})\,\;\quad m=1,2,\ldots, n^*\;,
\end{equation}
with $\eta_k(\ell_{0,m}) \overset{ind}{\sim} N(0, \gamma_{k,j}^2)$ and an NNGP predictive model for $u_k(\ell_{0,m})$. We briefly explain below and refer the reader to further details in \cite{Banerjee2017} Section~3.2 and their Equation~(19). 

It will be convenient, in what follows, to denote $\bK(\calA, \calB)$ for finite sets of locations $\calA$ and $\calB$ to be the matrix whose $(i,j)$-th element is evaluated from the covariance function $K(\ell_i, \ell_j)$, where $\ell_i$ and $\ell_j$ are the $i$-th entry in $\calA$ and $j$-th entry in $\calB$, respectively. We build a sequence of neighbor sets $N(\ell_{0,m})$ for each $m=1,2,\ldots,n^{\ast}$ that consist of a fixed number of nearest neighbors of $\ell_{0,m}$ from the ``past,'' where ``past'' refers to the $n_s$ locations already accounted for in ${\cal L}_s$ and the set of $\ell_{0,i}$'s for $i < m$. The distribution of $u_k(\ell_{0,m})$ is specified as  
\begin{equation}\label{eq:modx_pred_latent}
    u_k(\ell_{0,m}) = \sum_{\ell' \in N(\ell_{0,m})} a(\ell_{0,m}, \ell')u_k(\ell') + \omega_k(\ell_{0,m})\,\;\quad m=1,2,\ldots, n^*\;,
\end{equation}
where solving $\bK(N(\ell_{0,m}), N(\ell_{0,m}))\ba(\ell_{0,m}) = \bK(N(\ell_{0,m}), \ell_{0,m})$ for the vector $\ba(\ell_{0,m})$ renders the values of $a(\ell_{0,m}, \ell')$ in (\ref{eq:modx_pred_latent}), while $\omega_k(\ell_{0,m}) \sim N(0, \delta^2_{0,m})$ with 
\[
\delta^2_{0,m} = K(\ell_{0,m}, \ell_{0,m}) - \bK(\ell_{0,m}, N(\ell_{0,m}))\bK(N(\ell_{0,m}), N(\ell_{0,m}))^{-1}\bK(N(\ell_{0,m}), \ell_{0,m})\;.
\]    
Predictive inference is achieved by sampling from $[x_k(\ell_{0,m})\given \bx_k]$, where $\bx_{k}$ is the collection of observations for the $k$-th canopy structure variable. Since the joint posterior distribution of the hierarchical model extended over ${\cal L}_0$ is $[\bOmega_{x_k}, \bx_k^{\ast}, \bu_k^{\ast}\given \bx_k] \propto [\bOmega_{x_k} \given \bx_k] \times [\bu_k^{\ast} \given \bOmega_{x_k}]\times [\bx_k^* \given \bu_k^{\ast}, \bOmega_{x_k}]$, we can use the posterior samples from $\bOmega_{x_k}^{(l)} \sim [\bOmega_{x_k} \given \bx_k]$ for each $l=1,2,\ldots,L$ and each $\ell_{0,m} \in {\cal L}_0$ to draw $u_k^{(l)}(\ell_{0,m}) \sim [u_k(\ell_{0,m}) \given \bOmega_{x_k}^{(l)}]$ and $x_k^{(l)}(\ell_{0,m})\sim [x_k(\ell_{0,m})\given u_k^{(l)}(\ell_{0,m}), \bOmega_{x_k}^{(l)}]$. The resulting $x_k^{(l)}(\ell_{0,m})$ are the desired posterior predictive samples.  

Predictive inference for $y(\ell)$ over ${\cal L}_0$ proceeds analogously. We draw $y^{(l)}(\ell_{0,m})$ from $[y(\ell_{0,m}) \given \bOmega_{y}^{(l)}, \bx^{(l)}(\ell_{0,m})]$ by generating a value from (\ref{eq:mody}) for each sampled $\bOmega_{y}^{(l)}$ and $\bx^{(l)}(\ell_{0,m})$, where the latter is generated from its posterior predictive distribution as above. We model biomass over an area $D$ as a density, i.e., on a per unit area basis Mg/ha, which we denote by $y(D)$. The posterior distribution for $y(D)$ is evaluated by calculating $y^{(l)}(D) \approx \sum^{n^\ast}_{m=1}y^{(l)}(\ell_{0,m})/n^{\ast}$ for each sampled $y^{(l)}(\ell_{0,m})$. Similarly, posterior samples of total biomass are obtained from $y_{Tot.}^{(l)}(D) = |D|\sum^{n^\ast}_{m=1}y^{(l)}(\ell_{0,m})/n^{\ast}$, where $|D|$ is the area in ha. Samples from the posterior distribution of $y(D)$ and $y_{Tot.}(D)$ are then summarized to described any quantity of interest, e.g., posterior distribution mean, standard deviation, and 95\% credible interval. 

\subsection{Implementation}\label{implementation}
Methods developed in the preceding sections were programmed in C++ and used \texttt{openBLAS} \cite{zhang13} and Linear Algebra Package (LAPACK; www.netlib.org/lapack) for efficient matrix computations. \texttt{openBLAS} is an implementation of Basic Linear Algebra Subprograms (BLAS; www.netlib.org/blas) capable of exploiting multiple processors. Computing details about efficiently updating NNGP spatial random effects are provided in \cite{Finley2019}. MCMC sampler code for (\ref{eq:mody}) and (\ref{eq:modx}) are provided in the supplemental material along with simulated data used for testing the code and estimation procedures.

The computer used for subsequent analyses was running a linux operating system with a AMD Ryzen Threadripper 3990X 64-Core Processor (128 threads) with 264 GB of RAM.

\subsection{Submodels and model selection}\label{sec:analysis_submodels_validation}
Connecting Sections \ref{sec:data} and \ref{sec:models}, for model (\ref{eq:mody}) our analysis uses the NLCD strata to form $\mathds{1}$ and single G-LiHT canopy structure variable $x_{CH}$ to form $\bX$ and $\tilde{\bX}$ (i.e., $q = 4$ and $p = 1$). Model (\ref{eq:modx}) for $x_{CH}$ is similarly informed using the NLCD strata and a single variable, $v_{TC}$, to form $\bV$ and $\tilde{\bV}$ (i.e., $r = 1$). 

We consider the full model (\ref{eq:mody}) and four submodels. All submodels follow model (\ref{eq:mody}) notation and indexing unless noted otherwise.
\begin{align*}
    \text{Submodel 1:} &\;\; y(\ell) = \beta_0 + x_{CH}(\ell)\beta_{CH} + \epsilon(\ell) \text{, where } \epsilon(\ell) \iid N(0, \tau^2),\\
    \text{Submodel 2:} &\;\; y(\ell) = \beta_0 + x_{CH}(\ell)\beta_{CH} + \epsilon(\ell) \text{, where } \epsilon(\ell) \overset{ind}{\sim} N(0, \tau^2_j),\\ 
    \text{Submodel 3:} &\;\; y(\ell) = \beta_0 + \tilde{\beta}_{0,j} + x_{CH}(\ell)(\beta_{CH} + \tilde{\beta}_{CH,j}) + \epsilon(\ell) \text{, where } \epsilon(\ell) \iid N(0, \tau^2),\\
    \text{Submodel 4:} &\;\; y(\ell) = \beta_0 + \tilde{\beta}_{0,j} + x_{CH}(\ell)(\beta_{CH} + \tilde{\beta}_{CH,j}) + \epsilon(\ell) \text{, where } \epsilon(\ell) \overset{ind}{\sim} N(0, \tau^2_j),\\
    \text{Full model:} &\;\; y(\ell) = \beta_0 + \tilde{\beta}_{0,j} + x_{CH}(\ell)(\beta_{CH} + \tilde{\beta}_{CH,j}) + w(\ell) + \epsilon(\ell)\text{, where } \epsilon(\ell) \overset{ind}{\sim} N(0, \tau^2_j).
\end{align*}
To assess the contribution of G-LiHT derived canopy structure information, we also consider the full model without $x_{CH}$, i.e., $y(\ell) = \beta_0 + \tilde{\beta}_{0,j} + w(\ell) + \epsilon(\ell)$.

We consider several criterion for selecting the ``best'' model. The deviance information criterion \citep[DIC;][]{spiegelhalter2002} and widely applicable information criterion \citep[WAIC;][]{Watanabe2010} model fit criterion were computed for each candidate model. DIC equals $-2(\text{L}-\text{p}_D)$ where L is goodness of fit and p$_D$ is a model penalty term viewed as the effective number of parameters. Two WAIC criteria were computed based on the log pointwise predictive density (LPPD) with $\text{WAIC}_1 = -2(\text{LLPD} - \text{p}_1)$ and $\text{WAIC}_2 = -2(\text{LLPD} - \text{p}_2)$ where penalty terms p$_1$ and p$_2$ are defined in \cite{Gelman2014} just prior to, and in, their Equation~(11). Models with lower DIC, WAIC$_1$, and WAIC$_2$ values have better fit to the observed data and should yield better out-of-sample prediction, see \cite{Gelman2014}, \cite{Vehtari2017}, or \cite{Green2020} for more details. Lastly, we compute root mean squared error (RMSE) between the observed and model fitted values.

\section{Design-based and model-based estimates}\label{sec:design}
Design- and model-based approaches follow different theories of inference. Both are well developed and compared in statistical and forestry literature \citep[see, e.g.,][]{Sarndal1978, Sarndal2003, Gregoire1989, mcroberts2010probability}. \cite{little2004jasa} offers an excellent review from diverse perspectives. The design-based approach assumes a fixed finite population that can be accessible (in principle without error) through a census if all population units were observed. Randomness is incorporated via the selection of population units into a sample according to a randomized sampling design. A sampling design assigns a probability of selection to each sample. This is often effective when the variability and dependence across the population units can be adequately captured by the sampling design.

A comprehensive review and assessment of design-based and model-based inference for survey data is beyond the scope of this paper. Nevertheless, an increasingly prevalent view seems to be that if the units of the population exhibit associations or dependencies that are too complex to be accounted for by a sampling design, then a model-based approach to inference is preferable \citep[see, e.g., the developments in][for model-based as well as fully Bayesian perspectives to inference for finite populations]{little2004jasa, Ghosh:2012md, banerjee2023finite}. Furthermore, it is well-known that popular design-based estimators such as the Horvitz-Thompson \citep{Horvitz:tp} emerge as special cases of certain hierarchical models \citep[see, e.g., Section~2 in][]{banerjee2023finite}. While design-based inference has been developed in SAE problems \citep[see, e.g., the text by][]{rao2015small}, in applications such as here the nature of the spatial structure in the data and complex dependencies suggest that the population is a realization from a data-generating stochastic process. As argued in \cite{banerjee2023finite}, introducing the complex dependencies arising from a spatial process is difficult in design-based weights. Post-stratification issues for spatial data are also complicated. This renders the model-based approach more appropriate, where randomness is incorporated through probability laws endowed by a spatial process. 

Design-based estimates presented in Section~\ref{sec:results} are generated using the FIA post-stratified estimators detailed in \cite{bechtold2005enhanced}. Here, the mean biomass within the $j$-th stratum is estimated as
\begin{equation}
    \bar{y}_j = \frac{\sum^{n_j}_{i=1}y_{j,i}}{n_j}
\end{equation}
with associated standard error of the mean estimated as 
\begin{equation}
    s_{\bar{y}_j} = \sqrt{\frac{\sum^{n_j}_{i=1}\left(y_{i,j} - \bar{y}_j\right)^2}{n_j\left(n_j-1\right)}},
\end{equation}
where $n_j$ is the number of plots within the $j$-th stratum. The post-stratified estimator combines stratum estimates using area-based weights. The $j$-th stratum's weight is $W_j=A_j/A$, where $A_j$ is the $j$-th stratum area and $A$ is the area of all strata (i.e., $\sum^{q}_{j=1}A_j$). The post-stratified mean is estimated as
\begin{equation}
\bar{y} = \sum^{q}_{j=1}W_j\bar{y}_j
\end{equation}
with associated standard error of the mean estimated as 
\begin{equation}
s_{\bar{y}}= \sqrt{\frac{1}{n}\left(\sum^q_{j=1}W_jn_js^2_{\bar{y}_j} + \sum^q_{j=1}\left(1-W_j\right)\frac{n_h}{n}s^2_{\bar{y}_j}\right)}.
\end{equation}

The different uses of randomization distributions between design- and model-based inference yield different population parameter estimates, particularly for uncertainty summaries of these estimates. In our setting, for example, a design-based 95\% confidence interval for the TIU's biomass total is interpreted as ``if a large number of independent and equally sized samples were collected according to the sampling design, 95\% of the intervals computed using these samples would include the population total.'' In contrast, a Bayesian model-based 95\% credible interval is interpreted as ``there is a 95\% chance or probability the interval includes the population total, given the posited model and observed data.'' 

While the interpretation of these intervals is different, it is worth pointing out two important benefits of the model-based approach in our application. First, we are able to predict the biomass at arbitrary locations while accounting for its dependence on canopy structure and using the spatial random field associated with biomass. Design-based estimators, on the other hand, will not offer inference beyond the finite population units. Second, our data analysis reveals that exploiting the spatial structure and the use of a model for canopy structure delivers improved uncertainty quantification in the biomass for different strata. We defer to our specific discussion surrounding the results in Tables~\ref{tab:tiu-db-ests}~and~\ref{tab:tiu-mb-ests}.  

\section{Analyses}\label{sec:analyses}

In Section~\ref{sec:results}, we present the full model estimates of biomass density and total for the TIU. We also present the design-based post-stratified estimates generated using the complete sample of 1\,091 TIU plots. In addition to the entire TIU, we consider two illustrative small areas---Caribou-Poker Creeks Research Watershed (CPC) and Bonanza Creek Experimental Forest (BCEF). CPC and BCEF are located on Alaska state land and run as long-term ecological research sites by the University of Alaska, Fairbanks. The location and extent of these small areas are shown in Figure~\ref{fig:tiu-data}.

As illustrated in Figures~\ref{fig:cpcrw-data} and \ref{fig:bnz-data}, CPC and BCEF have a continuous forest inventory (CFI) plot network where each plot follows the FIA layout and measurement protocol. These inventory data were used to generate design-based post-stratified estimates that we compare with our model-based estimates. Importantly, the TIU data inform the model-based SAEs, not the CPC and BCEF CFI data. In this way, design-based CPC and BCEF estimates provide an independent assessment using data separate from those used to inform the model-based estimates. 

The candidate models were fit using $n$=880 spatially coinciding FIA and G-LiHT locations (Figure~\ref{fig:tiu-data}). Model (\ref{eq:modx}) parameter estimates and subsequent $x_{CH}$ predictions were informed using the $n_s$=61\,029 G-LiHT only locations (Figure~\ref{fig:tiu-data}). Estimates for TIU biomass density and total were based on samples from posterior predictive distributions at $n^\ast$=2\,165\,220 locations (see, e.g., Figure~\ref{fig:tiu-data-zoom}) laid out on a 250-by-250 (m) grid (one prediction location per 6.25 ha). This chosen grid is sufficiently dense to yield stable areal estimates and provide detailed maps (see Figure~\ref{fig:estStability} for the relationship between estimate stability and prediction grid density over the TIU). Given CPC's and BCEF's small spatial extent and desire to provide more detailed prediction maps, estimates were based on samples from posterior predictive distributions at locations on a 50-by-50 (m) grid. 

Three chains of each model's MCMC sampler were run for 25\,000 iterations. The full model's runtime for one chain was $\sim$2 hours, with all submodels having substantially shorter runtimes. Posterior inference for all model parameters and predictions was based on 1\,000 post-burn and thinned MCMC samples from each of three chains. Runtime to generate the 3\,000 posterior predictive samples over each of the TIU's $n^\ast$ locations was $\sim$10 hours.  

\section{Results}\label{sec:results}
\subsection{Model selection}
Candidate model fit criteria scores are given in Table~\ref{tab:candModFits}. As suggested by Section~\ref{sec:eda} EDA, Submodel 2's improved fit over Submodel 1 and 3 supports stratum specific residual variances (i.e., $\tau^2_j$ parameters). Submodel 4, which includes stratum-varying coefficients and residual variances, provides the best fit among the submodels. The full model, with its spatially varying intercept, provides marginally better fit over that of Submodel 4. Given the strong relationships between biomass and $x_{CH}$ seen in the EDA, it is not surprising to see substantially degraded fit when this canopy structure variable is not included in the full model (i.e., comparing full model fit criteria with and without $x_{CH}$). Given the full model (\ref{eq:mody}) with $x_{CH}$ provides the best fit, it is used for all subsequent inference.

\begin{table}[ht!]
\begin{center}
{\small
\begin{tabular}{lrrrrr|r}
 &  \multicolumn{4}{c}{Submodels (with $x_{CH}$)} & \multicolumn{2}{c}{Full model}\\
\cmidrule(lr){2-5}\cmidrule(lr){6-7}
Criteria  &  \multicolumn{1}{c}{1}& \multicolumn{1}{c}{2} & \multicolumn{1}{c}{3} & \multicolumn{1}{c}{4} & With $x_{CH}$& Without $x_{CH}$\\
\midrule
WAIC$_1$ & 7369.82 & 7100.65 & 7112.79 & 6921.25 & \textbf{6894.00} & 8105.62\\
WAIC$_2$ & 7369.94 & 7102.80 & 7115.22 & 6924.03 & \textbf{6914.94} & 8047.47\\
p$_1$ & 10.30 & 18.47 & 20.18 & 30.98 & 152.05 & 71.48\\
p$_2$ & 10.36 & 19.54 & 21.39 & 32.37 & 162.52 & 42.40\\
LPPD & -3674.61 & -3531.86 & -3536.22 & -3429.64 & -3294.95 & -3981.33\\
\midrule
DIC & 7362.51 & 7088.07 & 7099.19 & 6899.82 & \textbf{6864.26} & 8056.36\\
p$_D$ & 2.99 & 5.89 & 6.57 & 9.56 & 122.31 & 22.21\\
\midrule
RMSE & 16.84 & 15.97 & 14.41 & 13.68 & \textbf{12.98} & 26.90\\
\bottomrule
\end{tabular}
}
\caption{Fit criterion for candidate models. Scores that indicate the ``best'' model are bolded.  
}\label{tab:candModFits}
\end{center}
\end{table}

\subsection{Tanana Inventory Unit estimates}\label{sec:resluts_tiu}
Estimates for the full model's stratum-varying regression coefficients and residual variance parameters  are given in Table~\ref{tab:fullEsts}. The remainder of this model's parameter estimates are given in Table~\ref{tab:fullEstsProcess}. As expected, coefficients' sign and magnitude generally follow EDA stratum specific regression coefficient estimates in Table~\ref{tab:edaEsts} and show $x_{CH}$ explains a substantial portion of variability in biomass (i.e., 95\% credible interval bound for each slope coefficient does not include zero). Compared with EDA models, the full model's information pooling, via stratum and spatial random effects, yields smaller and more precise residual variance estimates (i.e., $\tau^2_j$ parameters). Spatial process parameter estimates in Table~\ref{tab:fullEstsProcess} show spatial random effects capture a relatively small amount of residual variation (i.e., $\sigma^2_w$ to $\tau^2_j$s ratios are small). Spatial decay parameter estimates suggest mean effective spatial range is 37.92 (km) with lower and upper 95\% credible intervals 25.39 and 115.22 (km), respectively (we define the ``effective spatial range'' as the distance at which the spatial correlation drops below 0.05, which equals $-\log(0.05)/\phi_w$). 

\begin{table}[ht!]
\begin{center}
{\small
\begin{tabular}{lccc}
 & $\beta_0 + \tilde{\beta}_{0,j}$ & $\beta_{CH} + \tilde{\beta}_{CH,j}$ & $\tau^2_j$\\
\cmidrule(lr){2-2}
\cmidrule(lr){3-3}
\cmidrule(lr){4-4}
Stratum & $_{\text{(L. 95\%)}}\;\;\text{Mean}\;\;_{\text{(U. 95\%)}} $& $_{\text{(L. 95\%)}}\;\;\text{Mean}\;\;_{\text{(U. 95\%)}} $& $_{\text{(L. 95\%)}}\;\;\text{Mean}\;\;_{\text{(U. 95\%)}} $ \\
\midrule
Conifer&$_{(\text{-}4.213)}\;\;\text{-}1.735\;\;_{(0.879)}$&$_{(12.219)}\;\;12.945\;\;_{(13.699)}$&$_{(257.461)}\;\;314.385\;\;_{(382.148)}$\\
 Deciduous&$_{(\text{-}5.392)}\;\;\text{-}2.116\;\;_{(1.118)}$&$_{(8.288)}\;\;8.896\;\;_{(9.493)}$&$_{(246.176)}\;\;352.329\;\;_{(547.472)}$\\
 Mixed&$_{(\text{-}4.824)}\;\;\text{-}1.945\;\;_{(0.784)}$&$_{(8.221)}\;\;9.009\;\;_{(9.770)}$&$_{(97.547)}\;\;167.053\;\;_{(294.068)}$\\
 Other&$_{(\text{-}3.194)}\;\;\text{-}1.943\;\;_{(\text{-}0.849)}$&$_{(7.291)}\;\;7.721\;\;_{(8.145)}$&$_{(46.736)}\;\;69.811\;\;_{(84.513)}$\\
 \bottomrule
\end{tabular}
}
\caption{Biomass model (\ref{eq:mody}) parameter estimates, with $j$ indexing stratum. Associated process parameter estimates are given in Table~\ref{tab:fullEstsProcess}.}\label{tab:fullEsts}
\end{center}
\end{table}

Parameter estimates for $x_{CH}$'s model (\ref{eq:modx}) are given in Tables~\ref{tab:xfullEsts} and \ref{tab:xfullEstsProcess}. Here, estimates of stratum-varying regression coefficients suggest that both strata and $v_{TC}$ explain a substantial portion of variability in $x_{CH}$. The model's spatial random effect explains a portion of residual variation (i.e., $\nu^2_{CH,u}$ to $\gamma^2_j$s ratios are larger); however, the effective spatial range for the spatial process is short, extending only about 5 (km). This is not too surprising given $v_{TC}$ explains a substantial portion of $x_{CH}$'s variability. 

\begin{figure}[!ht]
\centering
\subfloat[]{\includegraphics[height=6cm,trim={0cm 3cm 0cm 2.5cm},clip]{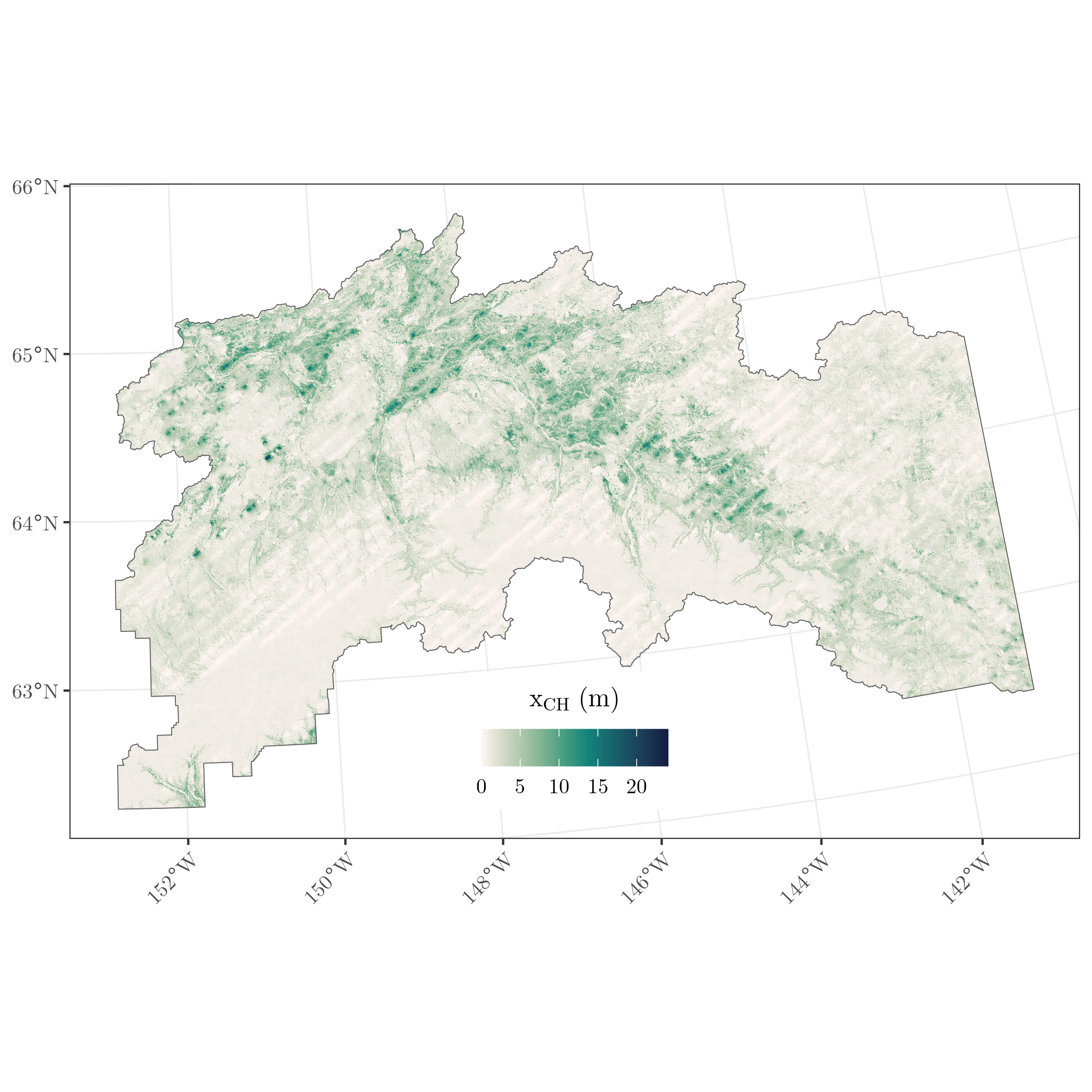}\label{fig:tiu-xmu}}
\subfloat[]{\includegraphics[height=6cm,trim={1cm 3cm 0cm 2.5cm},clip]{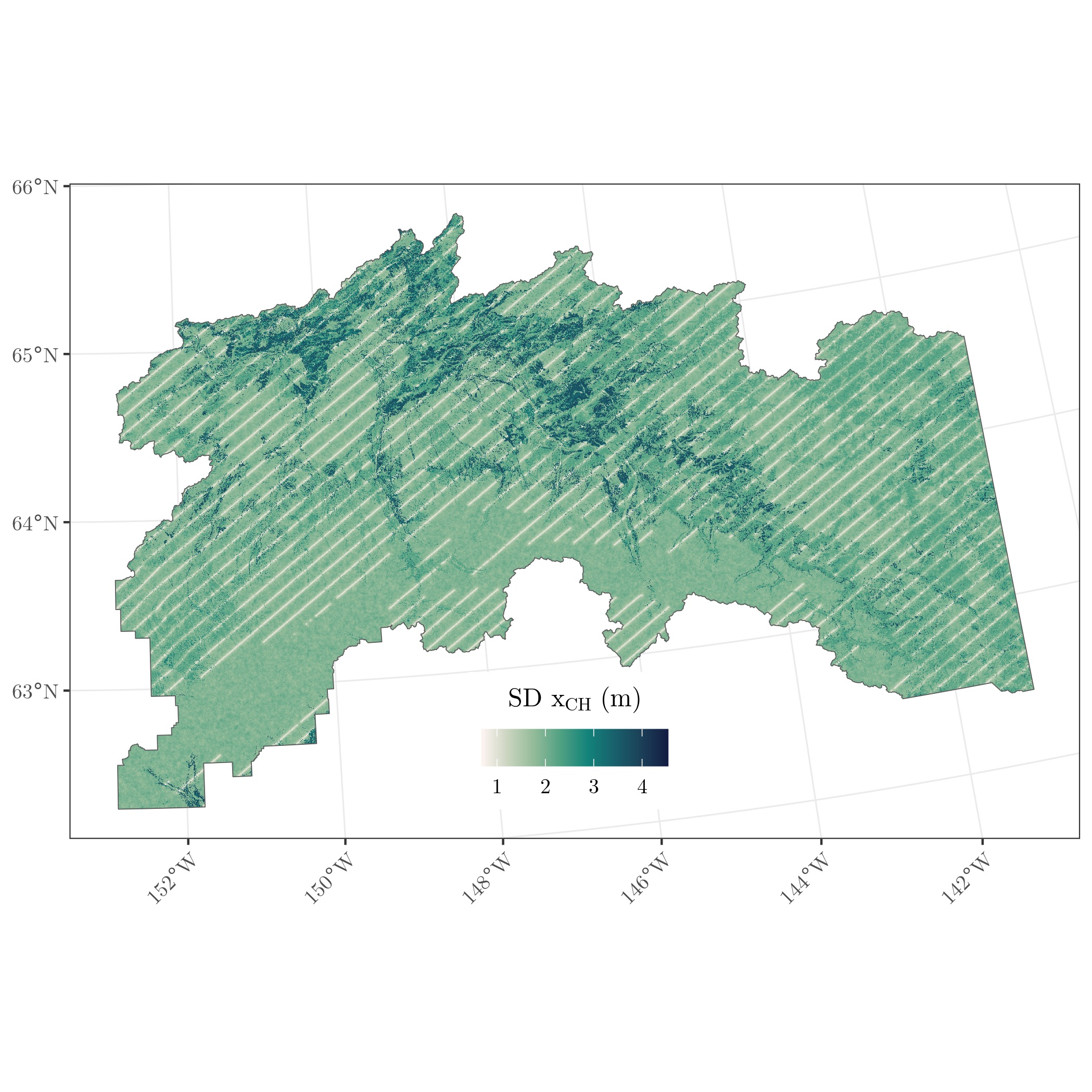}\label{fig:tiu-xsd}}
\caption{\protect\subref{fig:tiu-xmu} and \protect\subref{fig:tiu-xsd} mean and standard deviation, respectively, of the posterior predictive distribution for the G-LiHT mean canopy height variable estimated using (\ref{eq:modx}) and associated predictive model. }\label{fig:tiu-x}
\end{figure}

The predictive model for $x_{CH}$ (\ref{eq:modx_pred}) yields a posterior predictive distribution at the TIU's $n^*$ prediction locations. The posterior predictive distribution mean and standard deviation for $x_{CH}$ at each prediction location is given in Figures~\ref{fig:tiu-xmu} and \ref{fig:tiu-xsd}, respectively. Given the short effective spatial range, approximately 5 (km), the G-LiHT flight lines where $x_{CH}$ was measured appear as increased fidelity and precision stripes in the maps---predictions are improved along and adjacent to flight lines.

\begin{figure}[!ht]
\centering
\subfloat[]{\includegraphics[height=6cm,trim={0cm 3cm 0cm 2.5cm},clip]{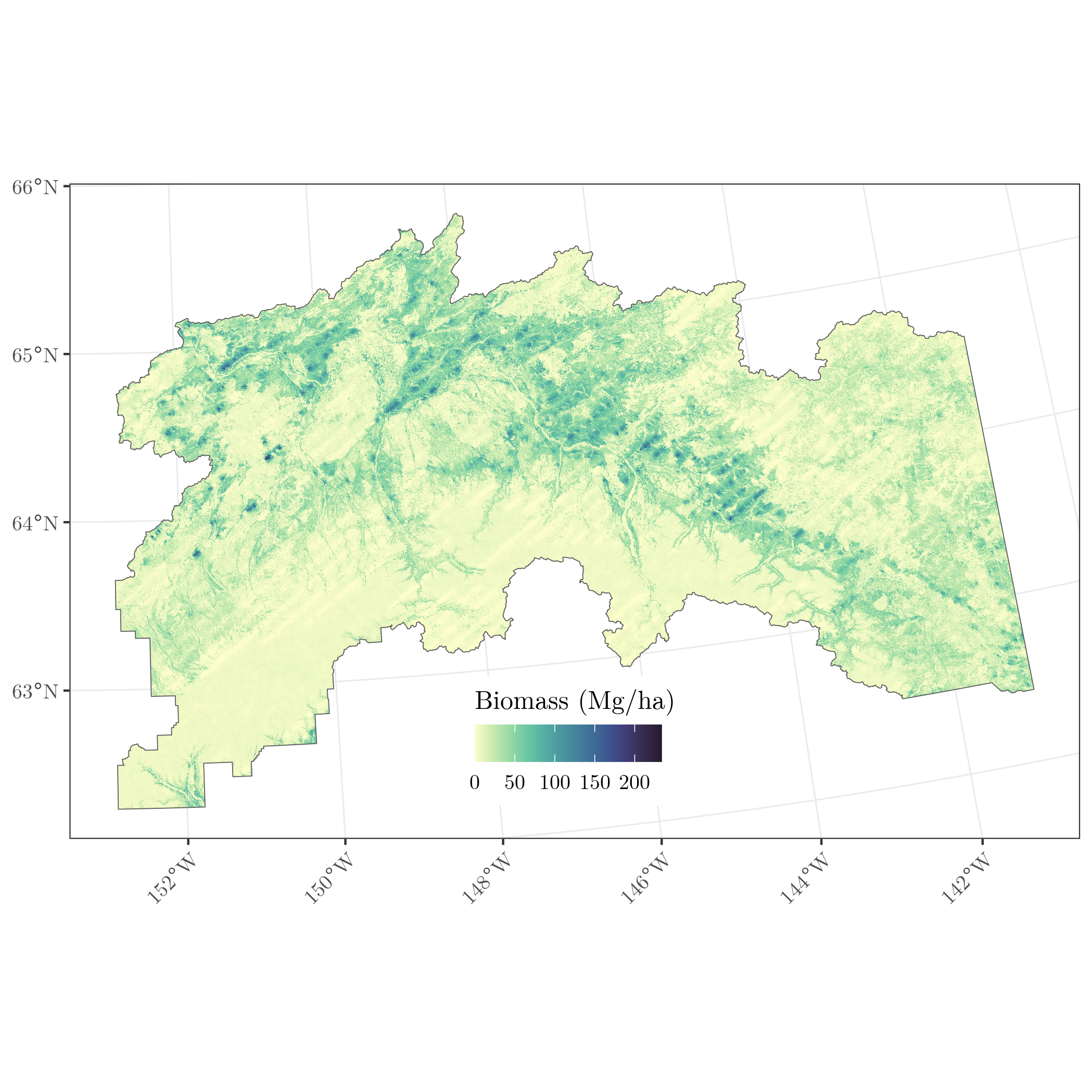}\label{fig:tiu-ymu}}
\subfloat[]{\includegraphics[height=6cm,trim={1cm 3cm 0cm 2.5cm},clip]{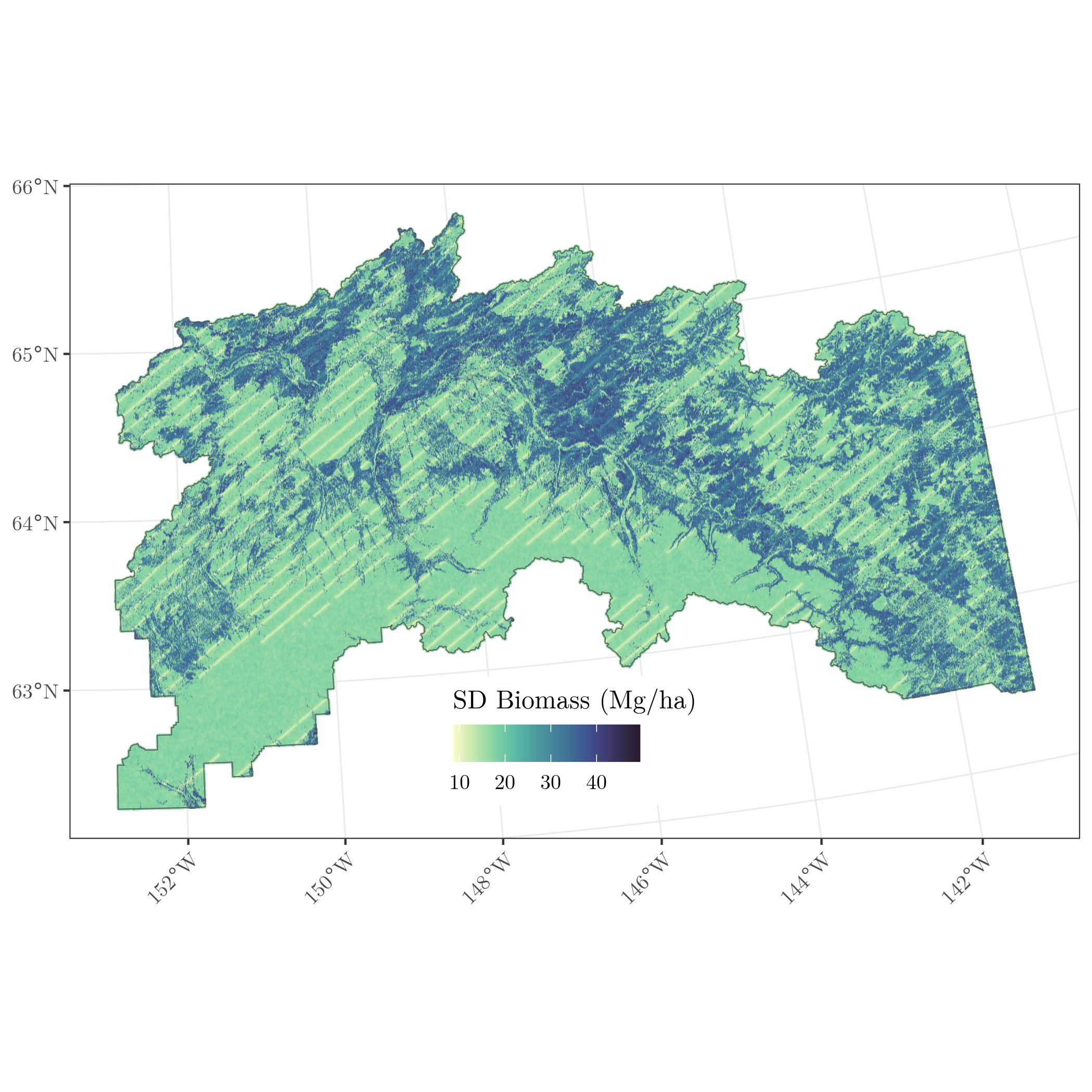}\label{fig:tiu-ysd}}
\caption{\protect\subref{fig:tiu-ymu} and \protect\subref{fig:tiu-ysd} mean and standard deviation, respectively, of the posterior predictive distribution for biomass estimated using (\ref{eq:mody}) and associated predictive model.}\label{fig:tiu-y}
\end{figure}

The two stage modeling approach developed in Section~\ref{sec:models} is designed to use $x_{CH}$'s posterior predictive information to inform biomass predictions. This information, propagated across model components, is seen as increased fidelity and precision along flight lines in biomass posterior predictive distribution summary maps Figures~\ref{fig:tiu-ymu} and \ref{fig:tiu-ysd}, respectively.

Turning now to area biomass estimates. Table~\ref{tab:tiu-db-ests} gives the stratum area, sample size, and design-based biomass density and total estimates. Corresponding model-based estimates derived from the $n^*$ posterior predictive distributions are given in Table~\ref{tab:tiu-mb-ests}. The estimators yield similar biomass density point estimates. The Mixed stratum shows the largest disparity, with a design-based estimate of 52.834 (Mg/ha) and model-based estimate of 40.288 (Mg/ha). 

The model-based point estimate for TIU biomass total is 9\,831.88 (1\,000 Mg) more than the design-based point estimate (i.e., $261\,441.680-251\,609.800 = 9\,831.88$). The Other stratum has the largest land area in the TIU. Hence, the seemingly small difference in biomass density between the design- and model-based estimates (i.e., 4.624 and 7.525, respectively) results in a large difference in biomass total (i.e., 39\,748.36 and 64\,689.31, respectively). Difference between the design- and model-based estimate for the Other stratum comprises about 62\% of the difference between the estimators' TIU biomass total estimate, i.e., 62\% of the 9\,831.88 (1\,000 Mg) is due to different Other density estimates (contribution of the remaining strata are Conifer 18\%, Deciduous 7\%, and Mixed 12\%).

Comparing stratum specific and TIU-wide design-based standard error to corresponding model-based posterior predictive distribution standard deviation in Tables~\ref{tab:tiu-db-ests} and \ref{tab:tiu-mb-ests} shows the model-based approach consistently yields more precise uncertainty quantification.

\begin{table}[ht!]
\begin{center}
\begin{tabular}{lrrrrrr}
\toprule
 &  &&\multicolumn{2}{c}{Biomass (Mg/ha)} & \multicolumn{2}{c}{Biomass (1000 Mg)}\\
\cmidrule(lr){4-5}\cmidrule(lr){6-7}
Stratum & Area (1000 ha) & $n$ &Mean & SE  & Total & SE \\
\midrule
Conifer&3659.234&278&36.183&2.240&132402.390&8197.627\\
 Deciduous&890.847&58&66.259&6.171&59026.840&5497.304\\
 Mixed&386.728&31&52.834&7.025&20432.210&2716.658\\
 Other&8596.150&724&4.624&0.542&39748.360&4655.796\\
\midrule
TIU&13532.960&1091&18.592&0.804&251609.800&10882.700\\
\bottomrule
\end{tabular}
\caption{Tanana Inventory Unit strata area, number of FIA plots $n$, and design-based estimates for biomass density and total with associated standard error (SE).}\label{tab:tiu-db-ests}
\end{center}
\end{table}

\begin{table}[ht!]
\begin{center}
\begin{tabular}{lcccc}
\toprule
 &  \multicolumn{2}{c}{Biomass (Mg/ha)} & \multicolumn{2}{c}{Biomass (1000 Mg)} \\
\cmidrule(lr){2-3} 
\cmidrule(lr){4-5} 
Stratum & $_{\text{(Lower 95\%)}}\;\;\text{Mean}\;\;_{\text{(Upper 95\%)}} $& SD & $_{\text{(Lower 95\%)}}\;\;\text{Mean}\;\;_{\text{(Upper 95\%)}} $& SD \\
\midrule
 Conifer&$_{(32.312)}\;\;34.188\;\;_{(36.502)}$&1.040&$_{(118238.780)}\;\;125102.840\;\;_{(133568.720)}$&3806.029\\
 Deciduous&$_{(58.907)}\;\;62.939\;\;_{(67.118)}$&2.179&$_{(52476.700)}\;\;56069.070\;\;_{(59791.490)}$&1941.281\\
 Mixed&$_{(35.632)}\;\;40.288\;\;_{(44.415)}$&2.163&$_{(13779.930)}\;\;15580.460\;\;_{(17176.670)}$&836.452\\
 Other&$_{(6.629)}\;\;7.525\;\;_{(8.571)}$&0.529&$_{(56981.520)}\;\;64689.310\;\;_{(73676.780)}$&4544.735\\
 \midrule
 TIU&$_{(18.312)}\;\;19.319\;\;_{(20.239)}$&0.513&$_{(247810.560)}\;\;261441.680\;\;_{(273898.360)}$&6945.950\\
\bottomrule
\end{tabular}
\caption{Tanana Inventory Unit model (\ref{eq:mody}) biomass density and total posterior predictive distribution mean and standard deviation (SD).}\label{tab:tiu-mb-ests}
\end{center}
\end{table} 
\pagebreak

\subsection{Small area estimates}
\begin{figure}[ht!]
\centering
\includegraphics[width=16cm]{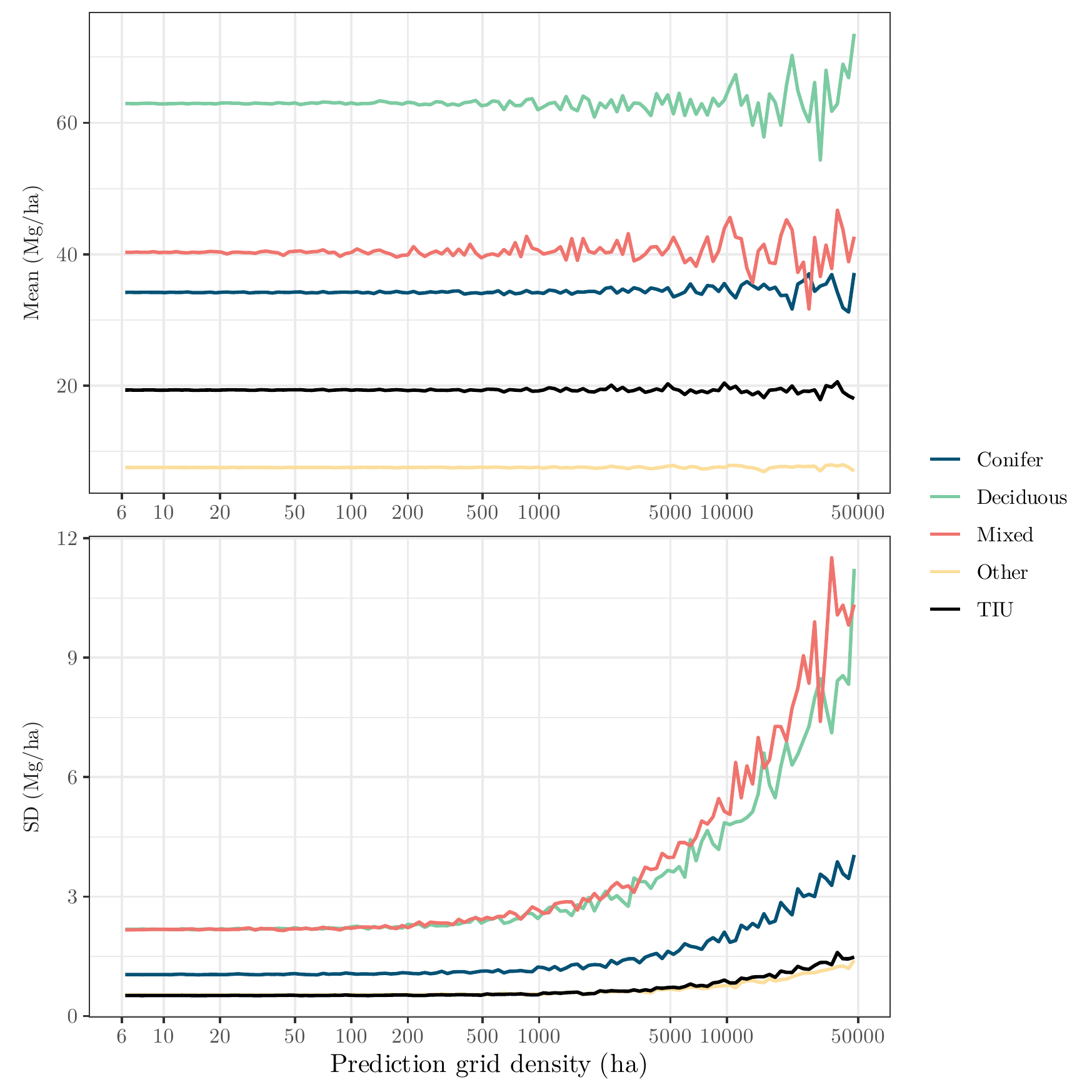}
\caption{Stratum specific and TIU-wide biomass posterior predictive distribution mean and standard deviation (SD) for a range of prediction grid densities. The x-axis ranges from one prediction location per 6.25 (ha) to one prediction location per 50\,000 (ha). Estimates in Table~\ref{tab:tiu-mb-ests} are based on one prediction location per 6.25 (ha). These figures show stratum specific mean and SD estimates are stable for densities of one prediction location per $\sim$50 (ha) and greater and hence support the choice of prediction grid density of one location per 6.25 (ha) for results reported in Section~\ref{sec:results}.}\label{fig:estStability}
\end{figure}

Next we consider the CPC and BCEF analysis results. As noted at the end of Section~\ref{sec:analysis_submodels_validation}, posterior inference comes directly from the TIU data and model; however, each small area uses a denser prediction location grid for posterior predictive inference and hence area estimates (i.e., one prediction every 0.25 ha for the small areas vs. 6.25 ha for the TIU). The denser grid yields more detailed maps and, given the fairly large size of each small area, has negligible effect on subsequent biomass density and total estimates (see, e.g., Figure~\ref{fig:estStability}).

\begin{figure}[!ht]
  \centering
  \begin{tabular}{@{}p{0.45\linewidth}@{\quad}p{0.45\linewidth}@{}}
    \subfloat{\subfigimg[width=\linewidth,trim={0cm 2.5cm 0cm 0.97cm},clip]{{\color{white}a)}}{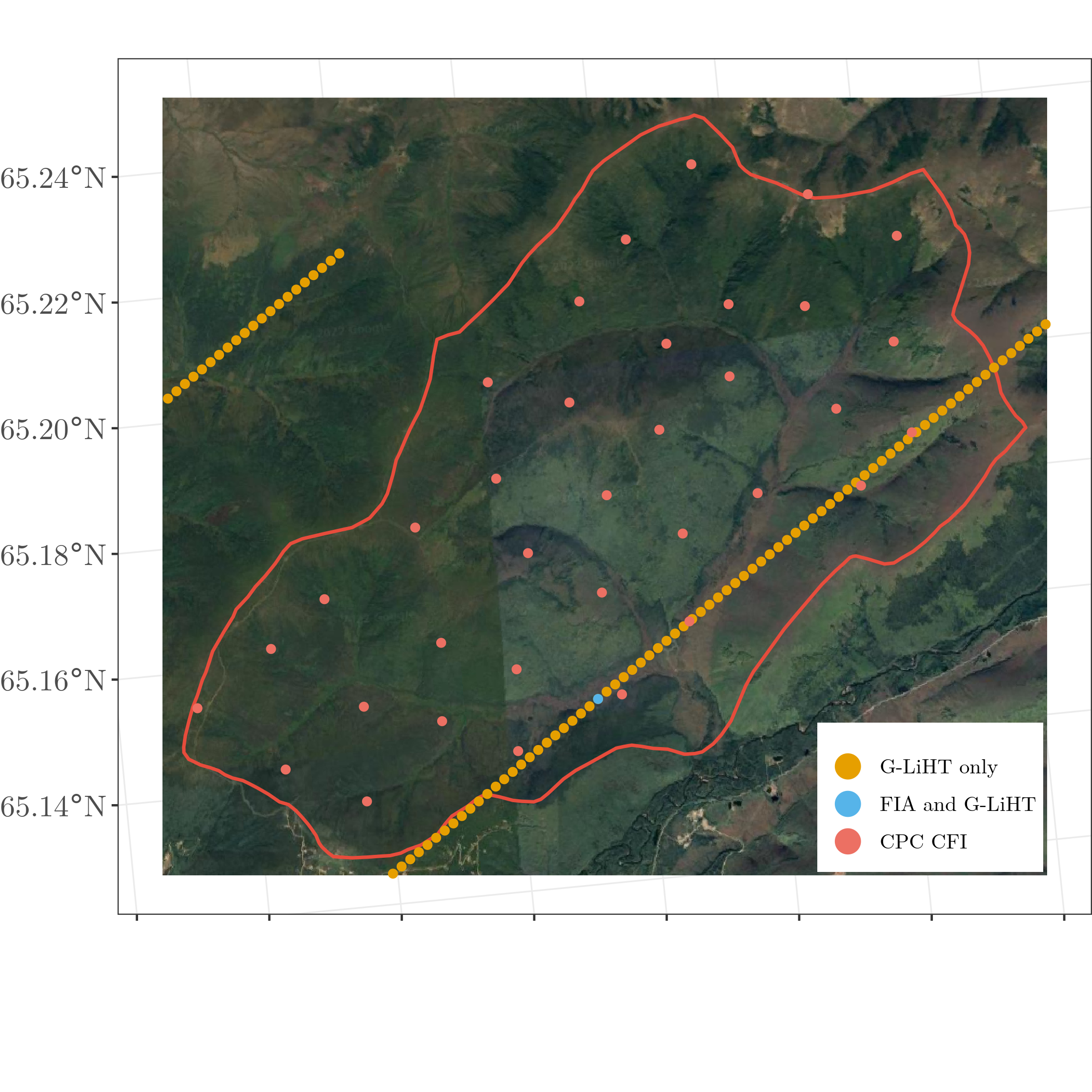}\label{fig:cpcrw-data}} &
    \subfloat{\subfigimg[width=\linewidth,trim={0cm 2.5cm 0cm 0.97cm},clip]{{\color{white}b)}}{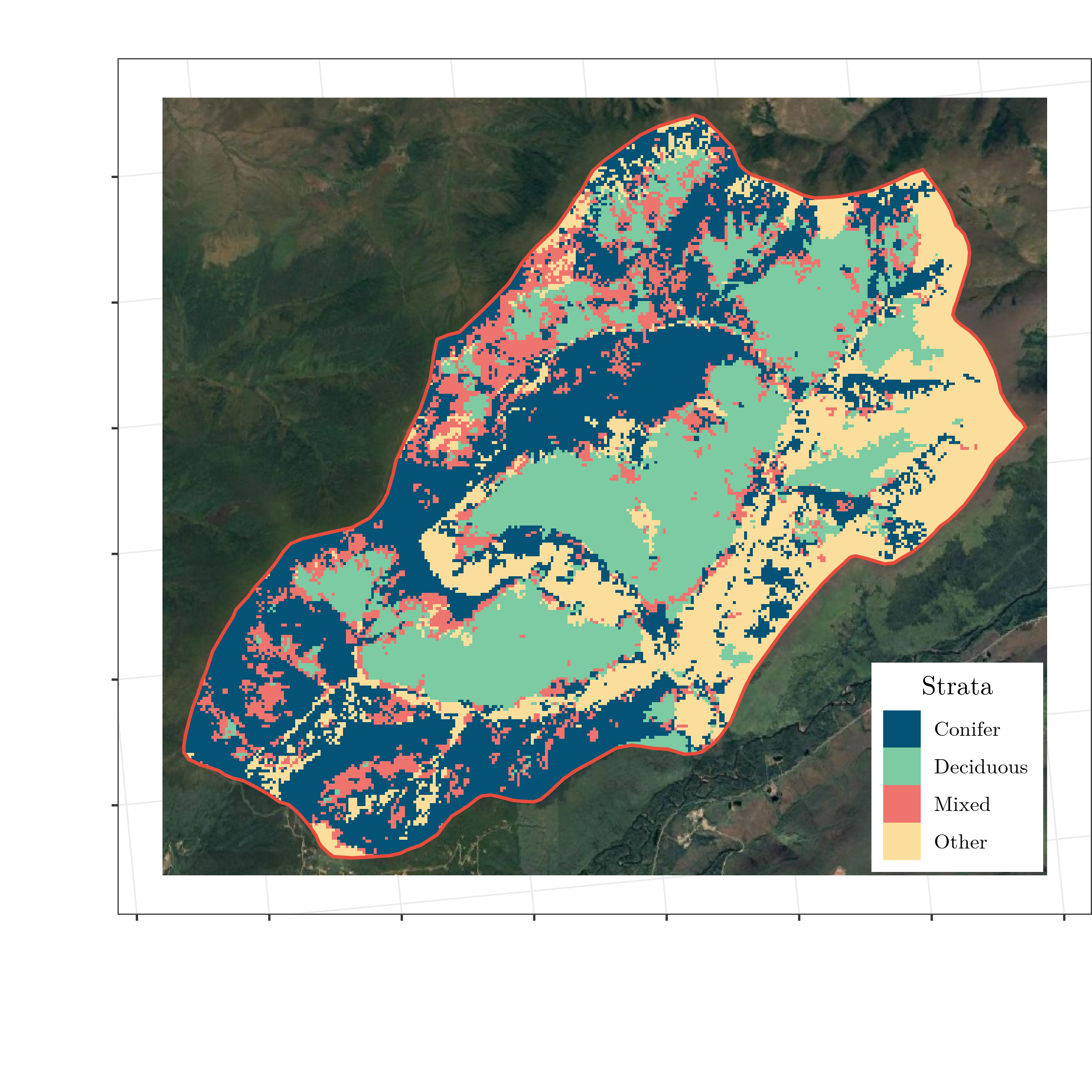}\label{fig:cpcrw-strata}} \\
    \subfloat{\subfigimg[width=\linewidth,trim={0cm 2.5cm 0cm 0.97cm},clip]{{\color{white}c)}}{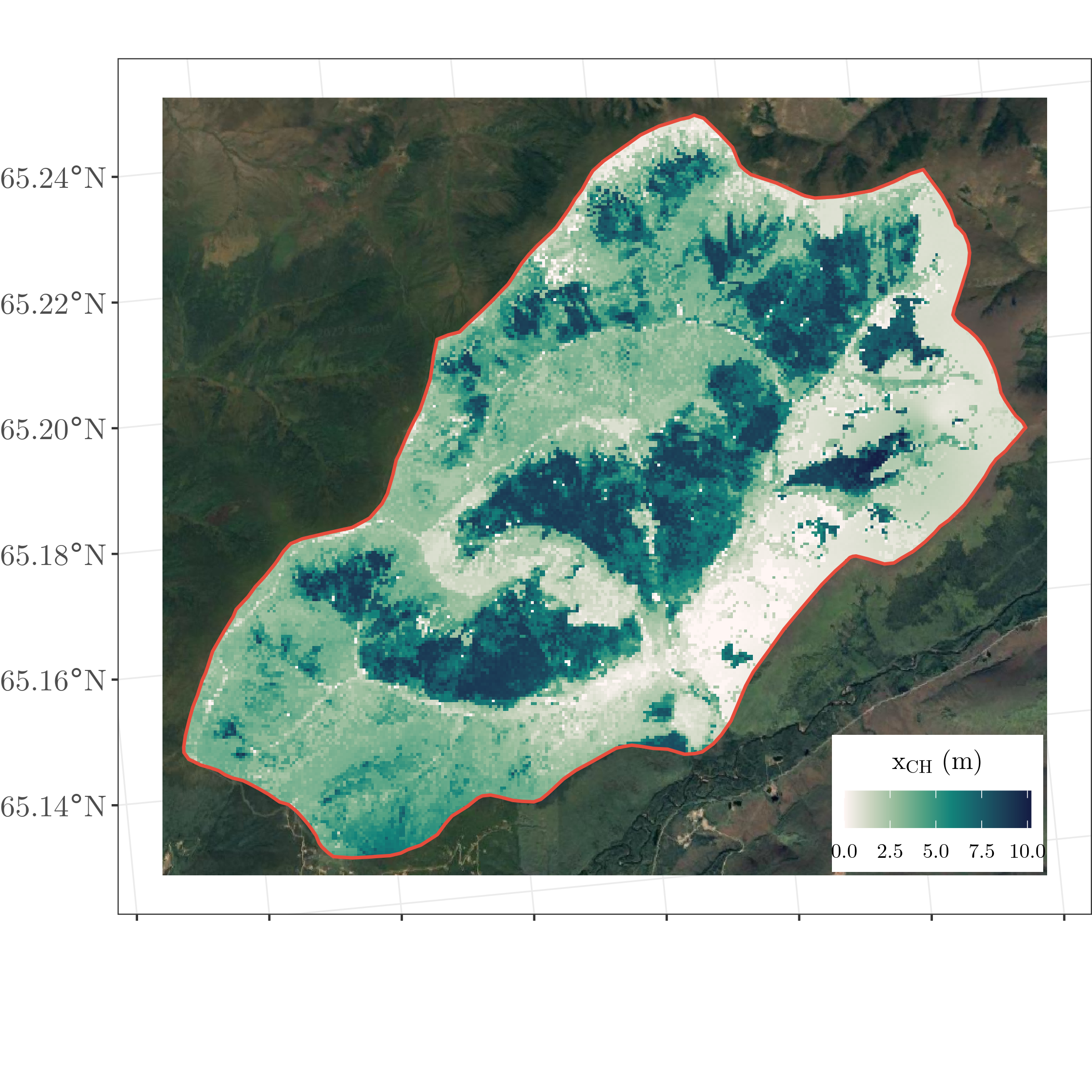}\label{fig:cpcrw-mux}} &
    \subfloat{\subfigimg[width=\linewidth,trim={0cm 2.5cm 0cm 0.97cm},clip]{{\color{white}d)}}{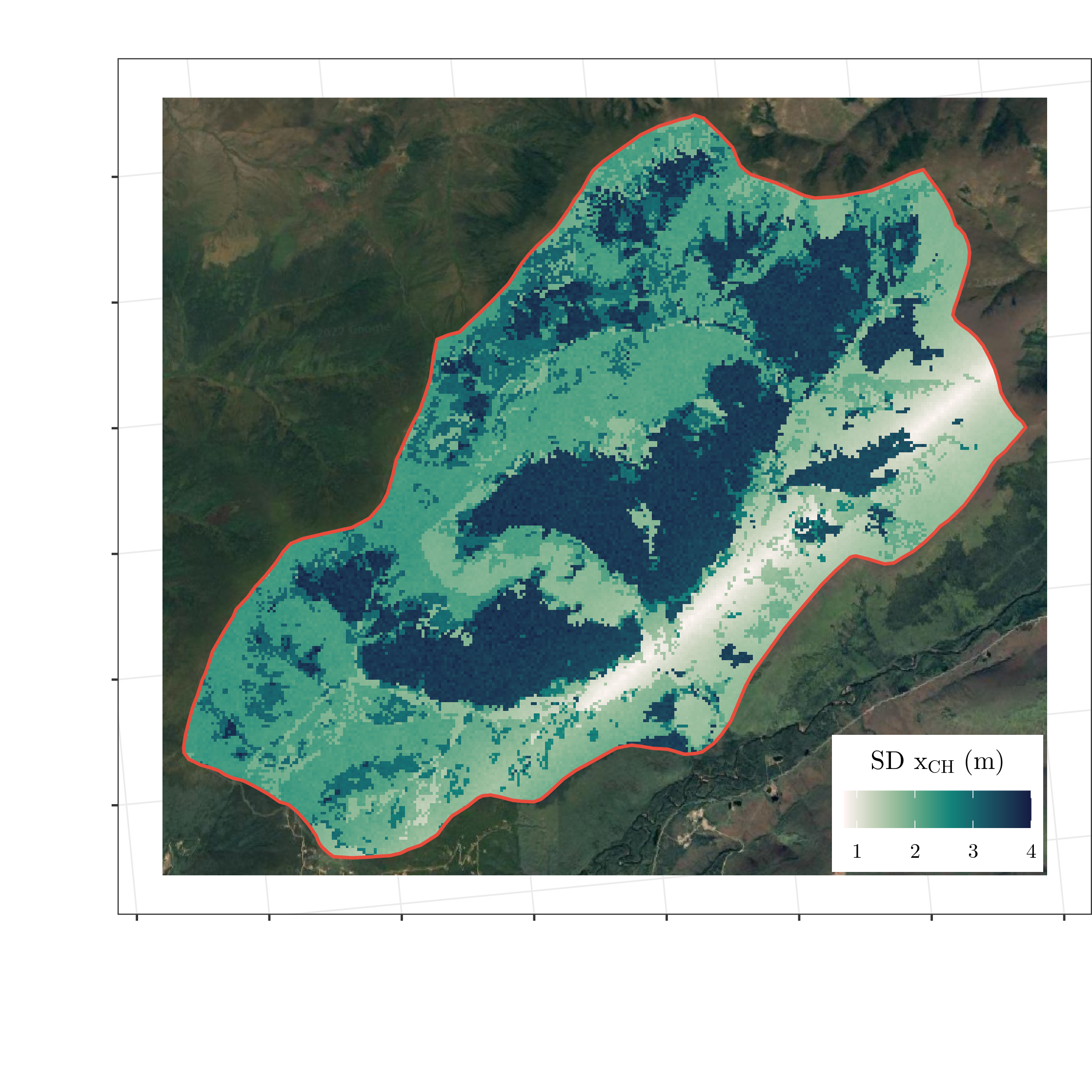}\label{fig:cpcrw-sdx}}  \\
    \subfloat{\subfigimg[width=\linewidth,trim={0cm 1cm 0cm 0.97cm},clip]{{\color{white}e)}}{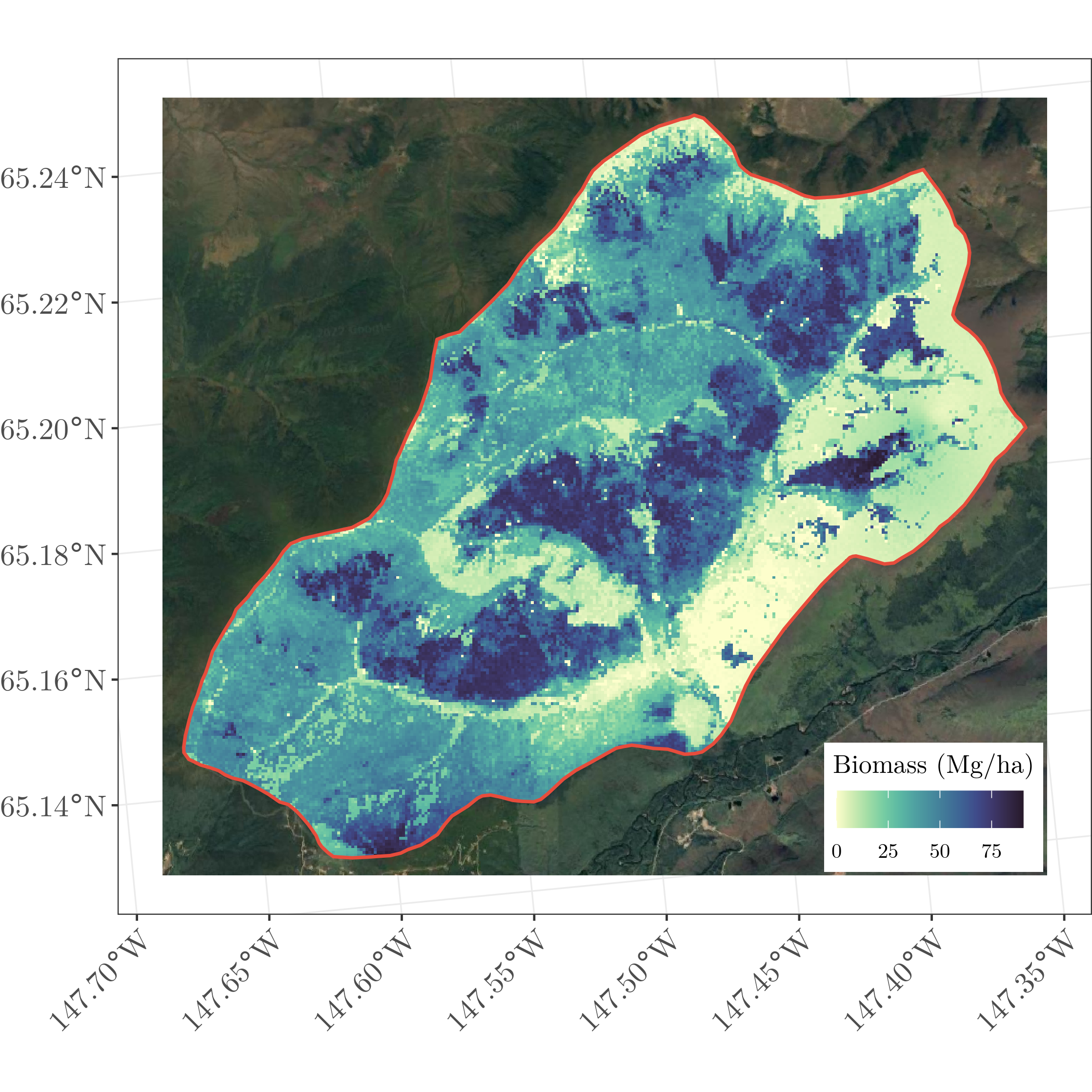}\label{fig:cpcrw-muy}} &
    \subfloat{\subfigimg[width=\linewidth,trim={0cm 1cm 0cm 0.97cm},clip]{{\color{white}f)}}{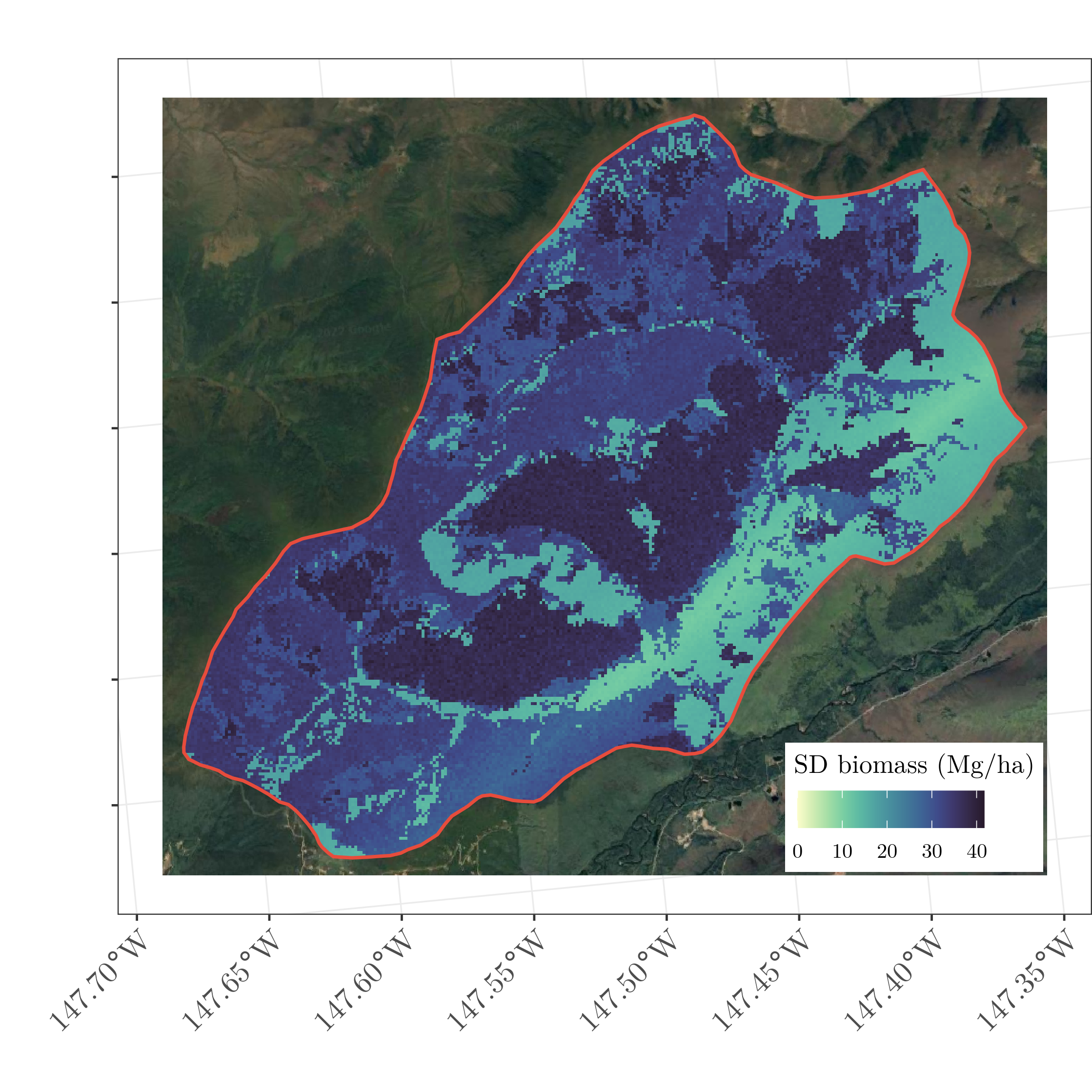}\label{fig:cpcrw-sdy}}
  \end{tabular}
  \caption{Caribou-Poker Creek Research Watershed (CPC) data and analysis results. \protect\subref{fig:cpcrw-data} locations of G-LiHT, FIA, and CPC's continuous forest inventory (CFI) data. \protect\subref{fig:cpcrw-strata} strata use for design- and model-based biomass estimates. \protect\subref{fig:cpcrw-mux} and \protect\subref{fig:cpcrw-sdx} mean and standard deviation, respectively, of the posterior predictive distribution for the G-LiHT mean canopy height variable estimated using (\ref{eq:modx}) and associated predictive model (\ref{eq:modx_pred}). \protect\subref{fig:cpcrw-muy} and \protect\subref{fig:cpcrw-sdy} mean and standard deviation, respectively, of the posterior predictive distribution for biomass estimated using (\ref{eq:mody}) and associated predictive model.} \label{fig:cp}
\end{figure}

We begin with the CPC, which is the northernmost small area shown in Figure~\ref{fig:tiu-data}. Figure~\ref{fig:cp}a shows the CPC boundary, locations with G-LiHT derived $x_{CH}$, single FIA plot, and CFI plot network. The strata are shown in Figure~\ref{fig:cp}b. 
The posterior predictive distribution mean and standard deviation for $x_{CH}$ at each prediction location is given in Figures~\ref{fig:cp}c and \ref{fig:cp}d, respectively. The increased precision seen in the broader TIU $x_{CH}$ standard deviation map Figure~\ref{fig:tiu-xsd} is less pronounced in Figure~\ref{fig:cp}d but still visible where the G-LiHT only observations coincide with the Other stratum (the large ratio of \emph{signal} variance $\nu^2_{CH,u}$ to \emph{noise} variance $\gamma^2_j$, where $j$ equals the Other stratum index, allows the random effect influence to be more apparent). Variability in $x_{CH}$ within a given stratum is due primarily to variability in tree cover $v_{TC}$ values. Biomass posterior predictive distribution mean and standard deviation maps are given in Figures~\ref{fig:cp}e and \ref{fig:cp}f, respectively. These maps make apparent the stratum-varying biomass density and variability, and propagated $x_{CH}$ uncertainty.

Table~\ref{tab:cpc-db} gives the CPC's stratum areas, CFI sample size, and design-based post-stratified biomass density and total estimates. Comparing the CPC and TIU design-based estimates (i.e., Tables~\ref{tab:tiu-db-ests} and \ref{tab:cpc-db}), we see the CPC's stratum specific biomass density differs substantially from that of the broader TIU. For example, CPC's Conifer and Mixed strata have about half the density of the TIU, i.e., 15.261 vs. 36.183 (Mg/ha) and 28.434 vs. 52.834 (Mg/ha), respectively. Whereas the CPC's Deciduous and Other strata have a greater density than the TIU, i.e., 73.195 vs. 66.259 (Mg/ha) and 12.263 vs. 4.625 (Mg/ha). CPC model-based estimates derived from the $n^*$ posterior predictive distributions are given in Table~\ref{tab:cpc-mb}. Relative to the CPC's design-based biomass density point estimates, the model-based estimates look more similar to the broader TIU. This is not surprising, because the model draws information from the entire TIU dataset to inform CPC biomass estimates. Despite the differences between the design- and model-based stratum specific density and total estimates seen when comparing Tables~\ref{tab:cpc-db} and \ref{tab:cpc-mb}, the CPC-wide densities and totals are quite similar, i.e., 32.284 vs. 37.695 (Mg/ha) and 342.341 vs. 399.725 (1\,000 Mg), respectively. 

Comparing the CPC design-based standard errors in Table~\ref{tab:cpc-db} with model-based standard deviations in Table~\ref{tab:cpc-mb} shows the model delivers slightly larger but comparable stratum specific and CPC-wide uncertainty quantification despite having only one FIA plot located within the CPC.

\begin{table}[ht!]
\begin{center}
\begin{tabular}{crrrrrr}
\toprule
 &  &&\multicolumn{2}{c}{Biomass (Mg/ha)} & \multicolumn{2}{c}{Biomass (1000 Mg)}\\
\cmidrule(lr){4-5}\cmidrule(lr){6-7}
Stratum & Area (ha) & $n$ &Mean & SE  & Total & SE \\
\midrule
Conifer&3793.290&13&15.261&5.378&57.888&20.400\\
 Deciduous&2922.567&10&73.195&6.398&213.918&18.698\\
 Mixed&1413.225&6&28.434&7.290&40.183&10.303\\
 Other&2475.102&6&12.263&5.669&30.352&14.032\\
\midrule
CPC&10604.180&35&32.284&3.217&342.341&34.118\\
\bottomrule
\end{tabular}
\caption{Caribou-Poker Creek Research Watershed strata area, number of inventory plots $n$, and design-based estimates for biomass density and total with associated standard error (SE).}\label{tab:cpc-db}
\end{center}
\end{table}

\begin{table}[ht!]
\begin{center}
\begin{tabular}{ccccc}
\toprule
 &  \multicolumn{2}{c}{Biomass (Mg/ha)} & \multicolumn{2}{c}{Biomass (1000 Mg)} \\
\cmidrule(lr){2-3} 
\cmidrule(lr){4-5} 
Stratum & $_{\text{(Lower 95\%)}}\;\;\text{Mean}\;\;_{\text{(Upper 95\%)}} $& SD & $_{\text{(Lower 95\%)}}\;\;\text{Mean}\;\;_{\text{(Upper 95\%)}} $& SD \\
\midrule
Conifer&$_{(22.810)}\;\;36.338\;\;_{(49.723)}$&7.125&$_{(86.524)}\;\;137.839\;\;_{(188.614)}$&27.026\\
 Deciduous&$_{(52.516)}\;\;64.087\;\;_{(75.373)}$&5.983&$_{(153.481)}\;\;187.300\;\;_{(220.281)}$&17.486\\
 Mixed&$_{(24.555)}\;\;36.604\;\;_{(48.815)}$&6.240&$_{(34.702)}\;\;51.730\;\;_{(68.986)}$&8.818\\
 Other&$_{(1.907)}\;\;9.247\;\;_{(16.340)}$&3.645&$_{(4.721)}\;\;22.886\;\;_{(40.444)}$&9.021\\
 \midrule
CPC&$_{(27.620)}\;\;37.695\;\;_{(47.904)}$&5.350&$_{(292.884)}\;\;399.725\;\;_{(507.978)}$&56.731\\
\bottomrule
\end{tabular}
\caption{Caribou-Poker Creek Research Watershed model (\ref{eq:mody}) biomass density and total posterior predictive distribution mean and standard deviation (SD).}\label{tab:cpc-mb}
\end{center}
\end{table}

\begin{figure}[!ht]
  \centering
  \begin{tabular}{@{}p{0.45\linewidth}@{\quad}p{0.45\linewidth}@{}}
    \subfloat{\subfigimg[width=\linewidth,trim={0cm 2.5cm 0cm 1.05cm},clip]{{\color{white}a)}}{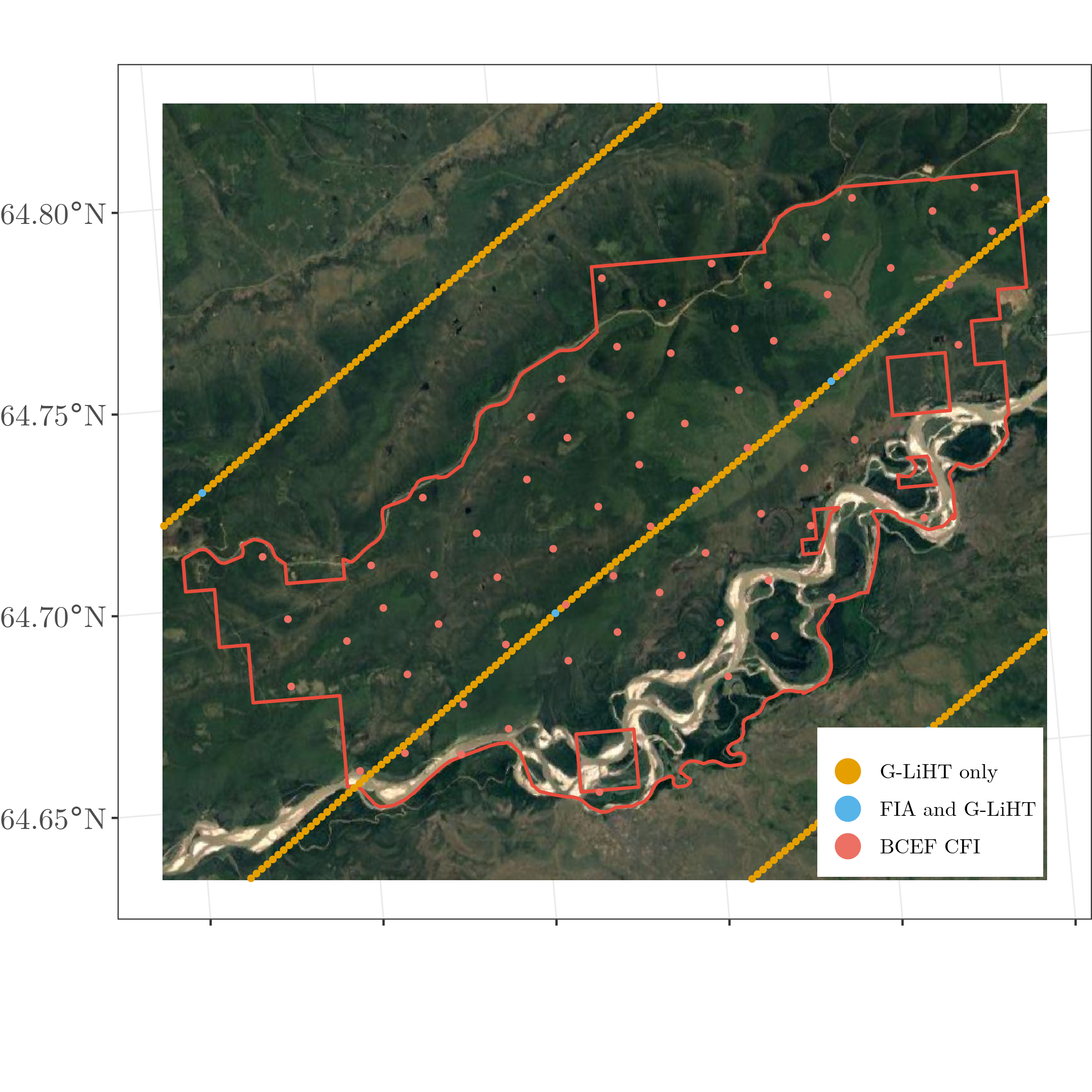}\label{fig:bnz-data}} &
    \subfloat{\subfigimg[width=\linewidth,trim={0cm 2.5cm 0cm 1.05cm},clip]{{\color{white}b)}}{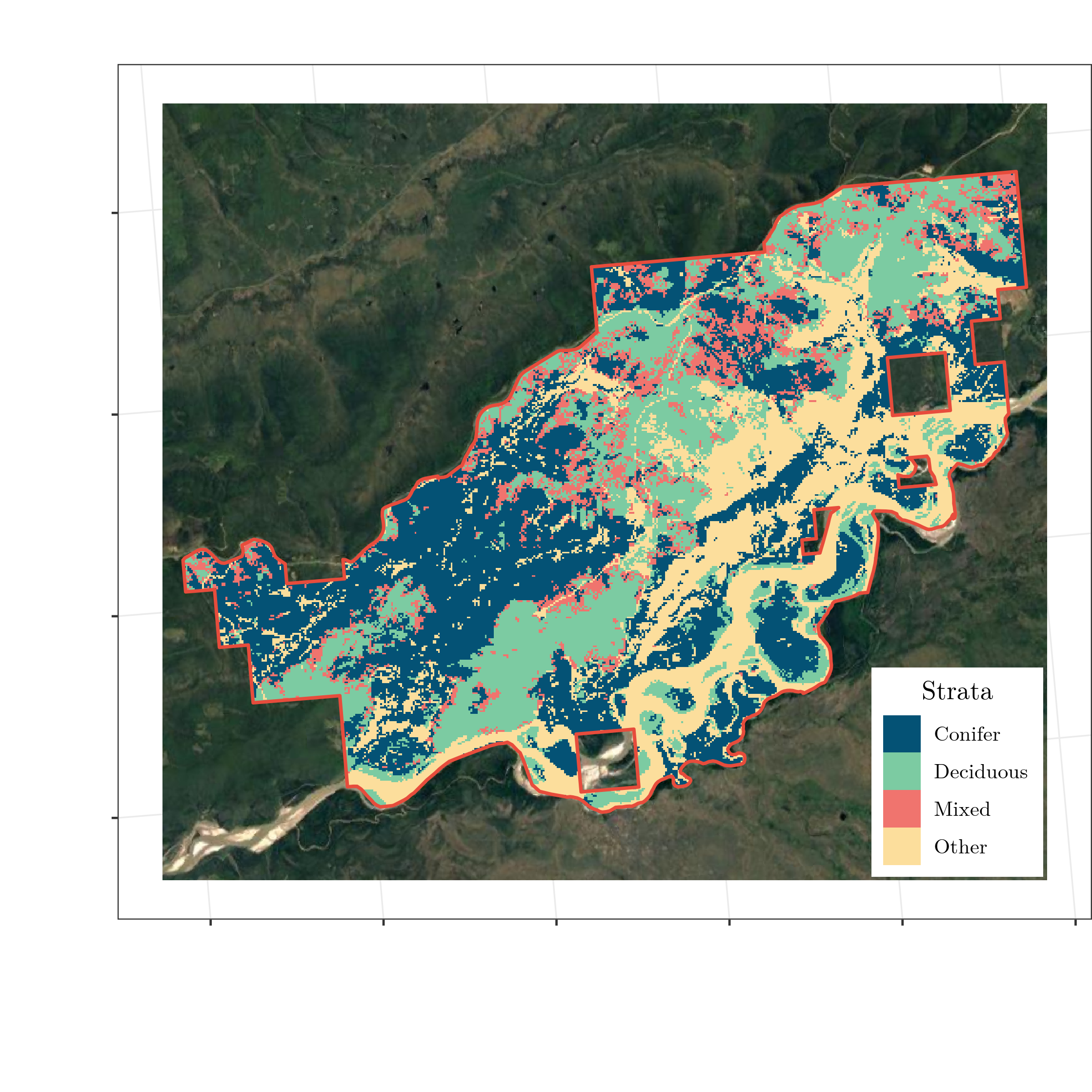}\label{fig:bnz-strata}} \\
    \subfloat{\subfigimg[width=\linewidth,trim={0cm 2.5cm 0cm 1.05cm},clip]{{\color{white}c)}}{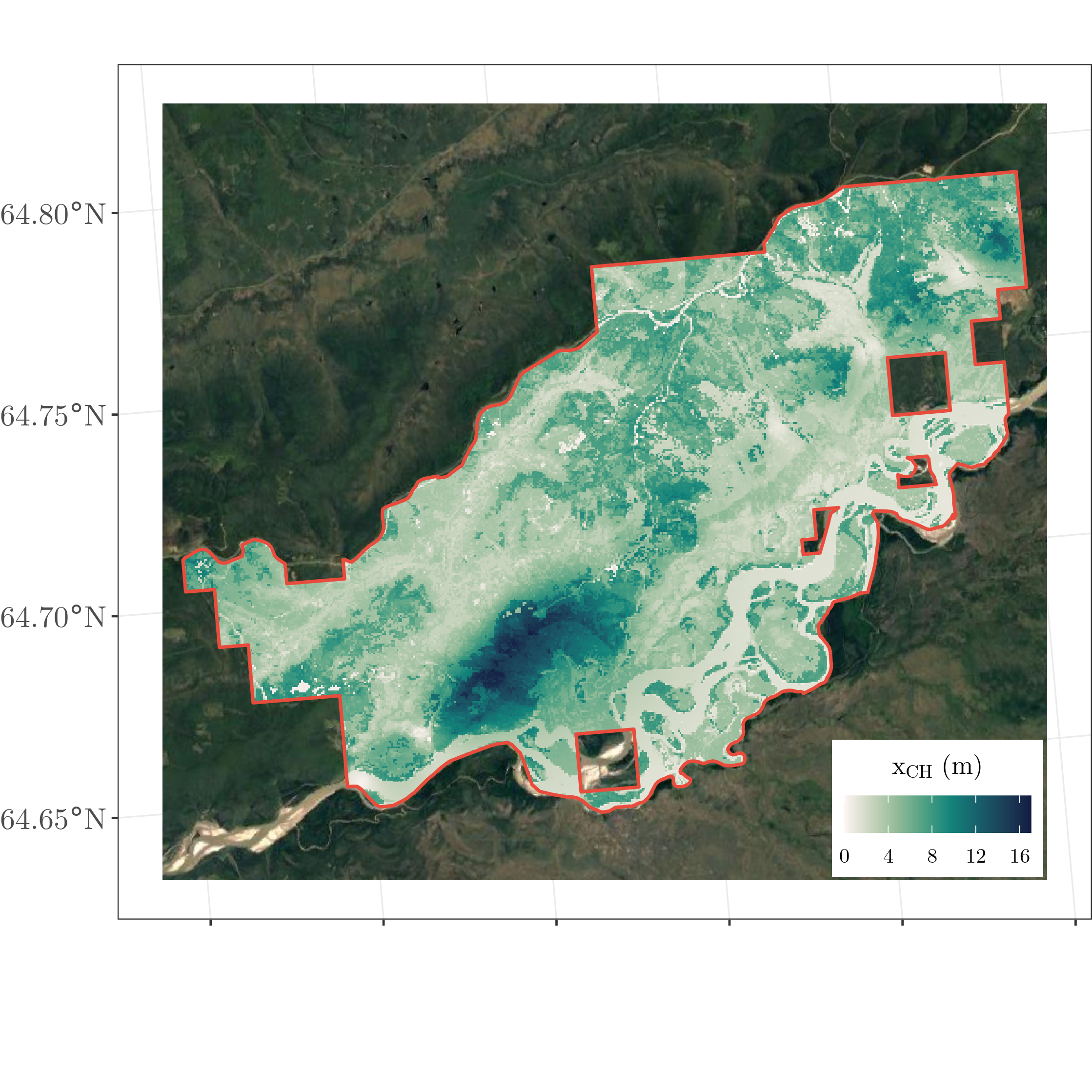}\label{fig:bnz-mux}} &
    \subfloat{\subfigimg[width=\linewidth,trim={0cm 2.5cm 0cm 1.05cm},clip]{{\color{white}d)}}{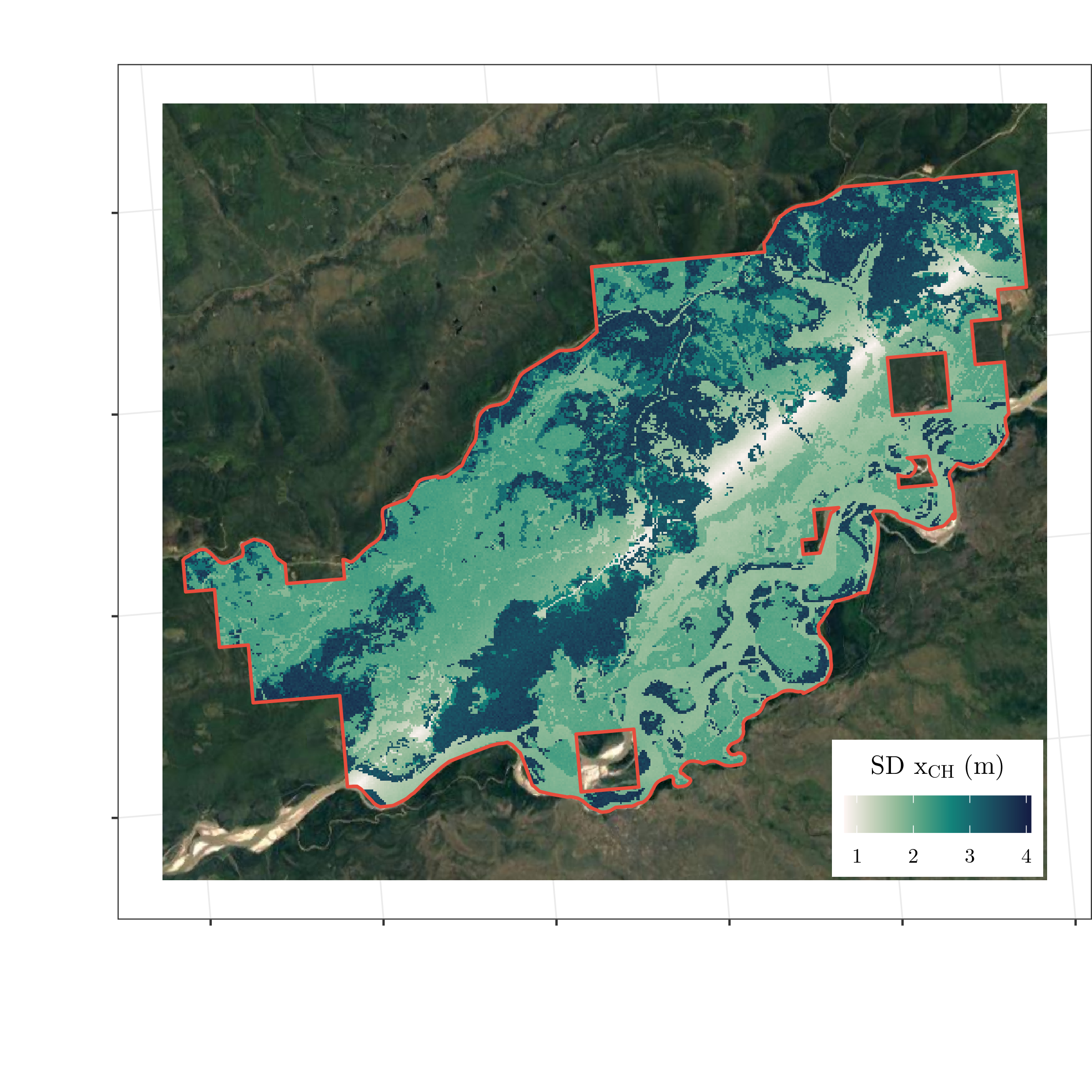}\label{fig:bnz-sdx}}  \\
    \subfloat{\subfigimg[width=\linewidth,trim={0cm 1cm 0cm 1.05cm},clip]{{\color{white}e)}}{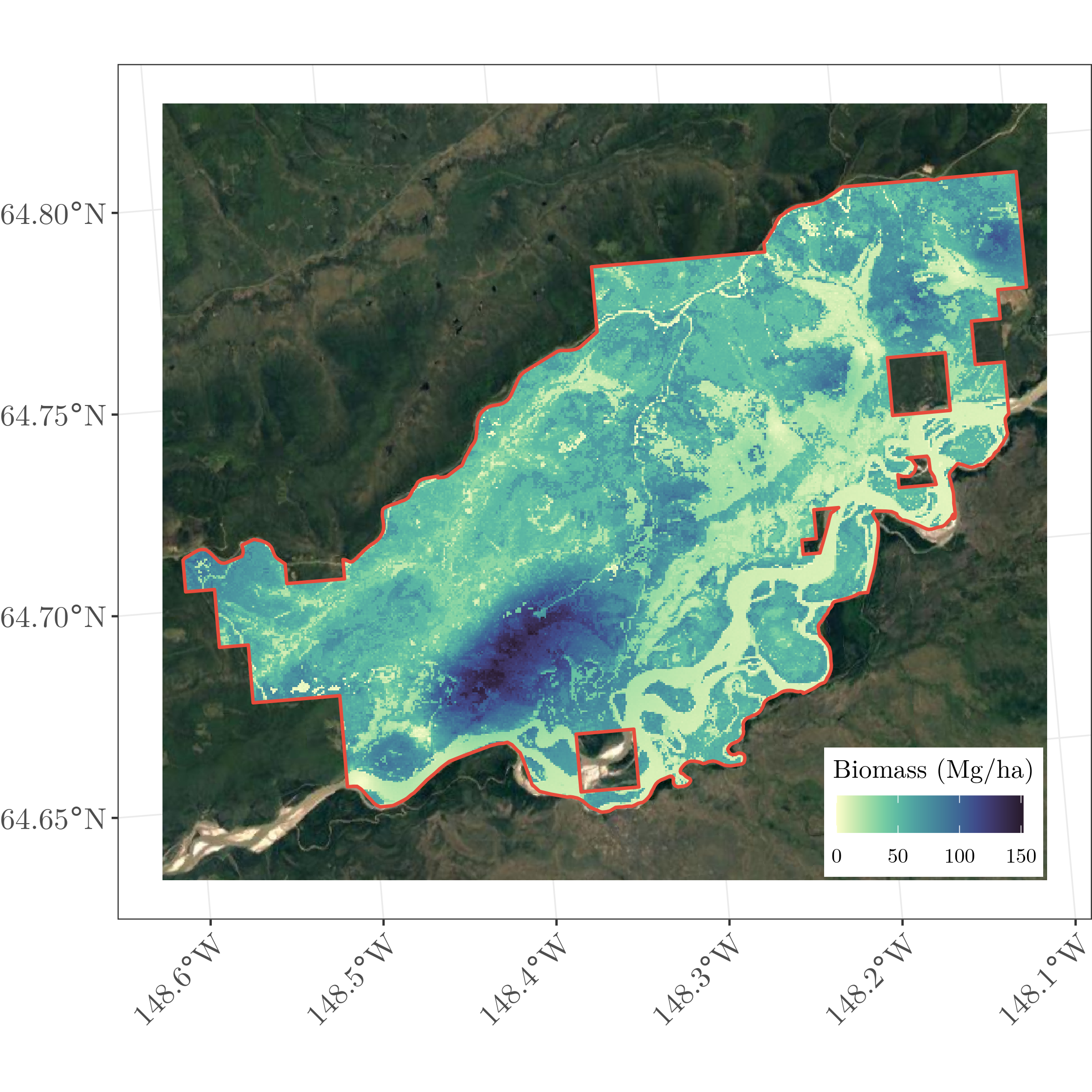}\label{fig:bnz-muy}} &
    \subfloat{\subfigimg[width=\linewidth,trim={0cm 1cm 0cm 1.05cm},clip]{{\color{white}f)}}{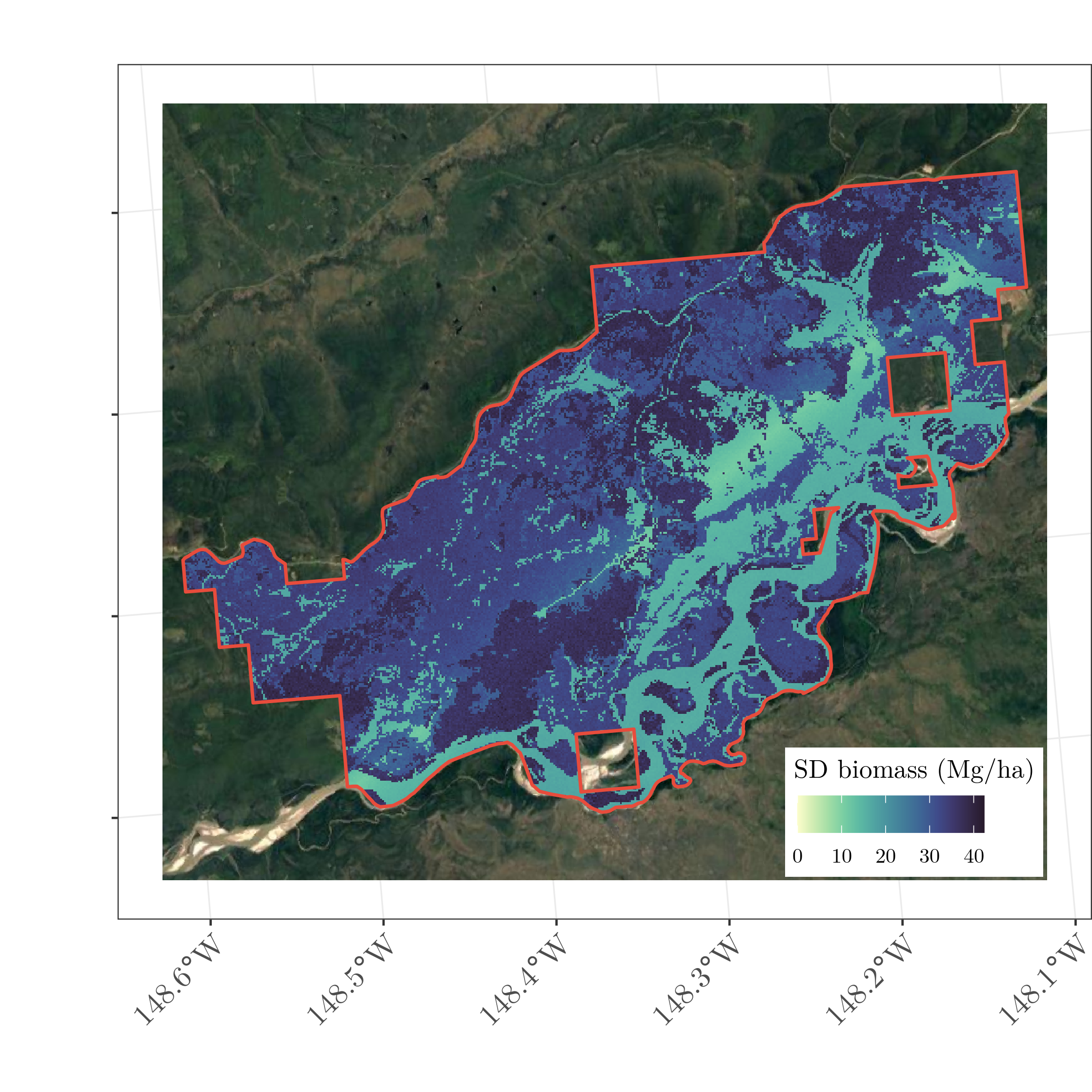}\label{fig:bnz-sdy}}
  \end{tabular}
  \caption{Bonanza Creek Experimental Forest (BCEF) data and analysis results. \protect\subref{fig:bnz-data} locations of G-LiHT, FIA, and BCEF's continuous forest inventory (CFI) data. \protect\subref{fig:bnz-strata} strata use for design- and model-based biomass estimates. \protect\subref{fig:bnz-mux} and \protect\subref{fig:bnz-sdx} mean and standard deviation, respectively, of the posterior predictive distribution for the G-LiHT mean canopy height variable estimated using (\ref{eq:modx}) and associated predictive model (\ref{eq:modx_pred}). \protect\subref{fig:bnz-muy} and \protect\subref{fig:bnz-sdy} mean and standard deviation, respectively, of the posterior predictive distribution for biomass estimated using (\ref{eq:mody}) and associated predictive model.} \label{fig:bnz}
\end{figure}

Next we consider BCEF analysis results. The BCEF is the southernmost small area shown in Figure~\ref{fig:tiu-data}. Figure~\ref{fig:bnz}a shows the BCEF boundary, locations with G-LiHT derived $x_{CH}$, two FIA plots, and CFI plot network. The strata are shown in Figure~\ref{fig:bnz}b. 
The posterior predictive distribution mean and standard deviation for $x_{CH}$ at each prediction location is given in Figures~\ref{fig:bnz}c and \ref{fig:bnz}d, respectively. Like the CPC, increased precision in $x_{CH}$ prediction (seen in Figure~\ref{fig:bnz}d) is most apparent in the Other stratum (due to the larger signal to noise ratio, i.e., $\nu^2_{CH,u}/\gamma^2_j$, where $j$ equals the Other stratum index). Biomass posterior predictive distribution mean and standard deviation maps are given in Figures~\ref{fig:bnz}e and \ref{fig:bnz}f, respectively.

Table~\ref{tab:bnz-db} gives the BCEF's stratum areas, CFI sample size, and design-based post-stratified biomass density and total estimates. The BCEF and TIU design-based biomass density estimates (i.e., Tables~\ref{tab:tiu-db-ests} and \ref{tab:bnz-db}) are similar, with the exception of the Other stratum. The Other point estimate for the BCEF is 18.070 vs. TIU's 4.624 (Mg/ha). BCEF's model-based estimates derived from the $n^*$ posterior predictive distributions are given in Table~\ref{tab:bnz-mb}. The BCEF's model-based biomass density point estimates for Conifer and Deciduous stratum are a bit higher than the design-based estimates. Despite differences between design- and model-based stratum specific density and total estimates seen when comparing Tables~\ref{tab:bnz-db} and \ref{tab:bnz-mb}, BCEF-wide densities and totals are quite similar, i.e., 41.397 vs. 49.391 (Mg/ha) and 867.503 vs. 1\,198.903 (1\,000 Mg), respectively. 

Comparing the BCEF design-based standard errors in Table~\ref{tab:bnz-db} with model-based standard deviations in Table~\ref{tab:bnz-mb} shows the model delivers consistently more precises uncertainty quantification across stratum specific and BCEF-wide estimates.

\begin{table}[ht!]
\begin{center}
\begin{tabular}{crrrrrr}
\toprule
 &  &&\multicolumn{2}{c}{Biomass (Mg/ha)} & \multicolumn{2}{c}{Biomass (1000 Mg)}\\
\cmidrule(lr){4-5}\cmidrule(lr){6-7}
Stratum & Area (1000 ha) & $n$ &Mean & SE  & Total & SE \\
\midrule
Conifer&7701.783&21&39.139&6.872&301.439&52.926\\
 Deciduous&5615.655&30&60.951&7.394&342.278&41.520\\
 Mixed&2124.105&8&58.444&17.532&124.140&37.239\\
 Other&5514.323&17&18.070&6.677&99.645&36.820\\
\midrule
BCEF&20955.870&76&41.397&4.157&867.503&87.106\\
\bottomrule
\end{tabular}
\caption{Bonanza Creek Experimental Forest strata area, number of inventory plots $n$, and design-based estimates for biomass density and total with associated standard error (SE).}\label{tab:bnz-db}
\end{center}
\end{table}

\begin{table}[ht!]
\begin{center}
\begin{tabular}{ccccc}
\toprule
 &  \multicolumn{2}{c}{Biomass (Mg/ha)} & \multicolumn{2}{c}{Biomass (1000 Mg)} \\
\cmidrule(lr){2-3} 
\cmidrule(lr){4-5} 
Stratum & $_{\text{(Lower 95\%)}}\;\;\text{Mean}\;\;_{\text{(Upper 95\%)}} $& SD & $_{\text{(Lower 95\%)}}\;\;\text{Mean}\;\;_{\text{(Upper 95\%)}} $& SD \\
\midrule
Conifer&$_{(43.211)}\;\;54.079\;\;_{(64.846)}$&5.396&$_{(332.800)}\;\;416.503\;\;_{(499.429)}$&41.557\\
 Deciduous&$_{(63.168)}\;\;71.335\;\;_{(80.474)}$&4.440&$_{(354.728)}\;\;400.591\;\;_{(451.914)}$&24.934\\
 Mixed&$_{(44.372)}\;\;54.328\;\;_{(63.964)}$&5.275&$_{(94.252)}\;\;115.397\;\;_{(135.866)}$&11.204\\
 Other&$_{(12.230)}\;\;18.600\;\;_{(25.275)}$&3.326&$_{(67.440)}\;\;102.569\;\;_{(139.372)}$&18.338\\
 \midrule
 BCEF&$_{(41.741)}\;\;49.391\;\;_{(57.211)}$&3.936&$_{(874.711)}\;\;1035.022\;\;_{(1198.903)}$&82.483\\
\bottomrule
\end{tabular}
\caption{Bonanza Creek Experimental Forest model (\ref{eq:mody}) biomass density and total posterior predictive distribution mean and standard deviation (SD).}\label{tab:bnz-mb}
\end{center}
\end{table}

\clearpage

\section{Discussion and future work}\label{sec:discussion}

We obtain stratum specific biomass estimates for arbitrarily sized areas, e.g., large area TIU and small area CPC and BCEF. From a SAE standpoint, the unit-level model we propose also provides maps for the given areas and at a user-defined spatial resolution. Under the Bayesian paradigm used here, these maps (and resulting areal estimates) can summarize a posterior predictive distribution characteristic, e.g., mean, median, standard deviation, or credible interval. Further, in the same way we condition biomass prediction on $x_{CH}$, access to biomass posterior predictive samples facilitates uncertainty propagation through other functions or models that take biomass as input (e.g., economic or ecological models).

Model assessment metrics used here suggest substantive information gain using G-LiHT derived mean canopy height, stratum-varying effects, a space-varying intercept, and stratum specific residual variance parameters. These features were identified using EDA. One might also consider allowing for stratum-varying spatial random effects \citep[see, e.g.,][]{Chan-Golston2020}; however, given the paucity of spatial structure after accounting for the mean canopy height, it is unlikely stratum-varying spatial processes are warranted.

Section~\ref{sec:analysis_submodels_validation} reveals that our model-based approach offers more precise uncertainty quantification for biomass than design-based inference (compare standard errors in Tables~\ref{tab:tiu-db-ests} with posterior standard deviations in Table~\ref{tab:tiu-mb-ests}) by exploiting the spatial structure in the data and incorporating information from canopy structure. With the acknowledgment that the true population parameter is unknown (and perhaps unknowable), we compare population point estimates. The design-based post-stratified point estimate, using 1\,091 FIA plots, puts TIU's total biomass at 251\,609.8 (1\,000 Mg). The model-based point estimate, using 880 FIA plots and auxiliary information, puts TIU's total biomass at a slightly higher 261\,441.68 (1\,000 Mg). Much of the difference in these estimates is due to the model placing more biomass in the Other stratum. Looking at the observed distribution of biomass in Other (Figure~\ref{fig:boxplots-bio}) and canopy cover map (Figure~\ref{fig:tiu-hansen}), future work should consider a refined set of strata that partition Other into non-vegetation (e.g., water, barren) and vegetation. Or, following \cite{Finley2011} and \cite{May2024}, one might introduce an additional hierarchical level to differentiate between locations with and without biomass.

We demonstrate the model-based approach is particularly useful for SAE. Here, paucity of FIA plots in CPC and BCEF precludes design-based estimation. In comparison, the proposed model draws on the broader TIU dataset, to deliver CPC and BCEF high-resolution maps and areal estimates by stratum. A key contribution here is demonstrating how a two-stage hierarchical Bayesian model allows information to be shared between multiple model components and how sparse data is used to inform parameter estimates within and across components, and ultimately to prediction. This model uses information where available to improve prediction accuracy and precision. For example, biomass prediction near G-LiHT measurements is seen in Figures~\ref{fig:tiu-ymu} and \ref{fig:tiu-xsd} as stripes of higher fidelity and increased precision, respectively. Away from G-LiHT flight lines, biomass prediction retreats to the stratum mean and its variance is dominated by that of the non-spatial residual process.

Our model is trained by the observed plots where both biomass and G-LiHT measurements are reported. We anticipate future data collection efforts in interior Alaska (beyond the TIU) will also align G-LiHT collection with all inventory plots. 
Since model-based inference is not limited to probability sampling, future work could explore adaptive sampling designs given objective functions that balance data acquisition cost at each model level (e.g., LiDAR and plot measurements) with inferential objectives (e.g., maximizing prediction accuracy and precision) \citep[see, e.g.,][]{Gangqiang2006, mateu2012spatio}. Given a posited model, like the one presented here, such adaptive sampling designs could facilitate cost-effective data collection efforts in remote regions while meeting inferential objectives. 

Finally, we anticipate substantial opportunities for expanding our modeling framework in future research. A natural extension would be to build multivariate process models \citep[see, e.g., Ch.9 in][]{banerjee2014hierarchical} that will jointly model biomass and an ensemble of data on canopy structures. These models, while offering richness in modeling through the use of cross-covariance functions or linear combinations of latent processes \citep{gelban10}, will need to reckon with a possibly large dimension of the response vector comprising all measurements on biomass and all of the canopy structures being considered. Here, the expanding literature on multivariate spatial factor models on very large domains, including for spatially misaligned data, offer exciting possibilities for further investigation \citep{ren2013hierarchical, TaylorRodriguez2019, zhang:2022sf}.


\section*{Acknowledgments}
Funding was provided by: NASA Carbon Monitoring System (CMS) grants Hayes (CMS 2020) and Cook (CMS 2018); National Science Foundation (NSF) DMS-1916395; joint venture agreement with the USDA Forest Service Forest Inventory and Analysis 22-CA-11221638-201; USDA Forest Service, Region 9, Forest Health Protection, Northern Research Station; Michigan State University AgBioResearch.
\newpage
\section*{Supplemental Material}
\beginsupplement

\begin{table}[ht!]
\begin{center}
{\small
\begin{tabular}{lc}
Parameter & $_{\text{(L. 95\%)}}\;\;\text{Mean}\;\;_{\text{(U. 95\%)}}$\\
\midrule
$\sigma^2_0$&$_{(0.332)}\;\;1.029\;\;_{(4.767)}$\\
 $\sigma^2_{CH}$&$_{(1.181)}\;\;3.124\;\;_{(10.130)}$\\
 $\phi_w$&$_{(0.026)}\;\;0.079\;\;_{(0.118)}$\\
 $\sigma^2_w$&$_{(6.767)}\;\;16.326\;\;_{(41.274)}$\\
 \bottomrule
\end{tabular}
}
\caption{Process parameter estimates for the full model with $x_{CH}$.  Biomass model (\ref{eq:mody}) process parameter estimates. Associated parameter estimates are given in Table~\ref{tab:fullEsts}.}\label{tab:fullEstsProcess}
\end{center}
\end{table}

\begin{table}[ht!]
\begin{center}
{\small
\begin{tabular}{lccc}
 & $\alpha_{CH,0} + \tilde{\alpha}_{CH,0,j}$ & $\alpha_{CH,TC} + \tilde{\alpha}_{CH,TC,j}$ & $\gamma^2_j$\\
\cmidrule(lr){2-2}
\cmidrule(lr){3-3}
\cmidrule(lr){4-4}
Stratum & $_{\text{(L. 95\%)}}\;\;\text{Mean}\;\;_{\text{(U. 95\%)}} $& $_{\text{(L. 95\%)}}\;\;\text{Mean}\;\;_{\text{(U. 95\%)}} $& $_{\text{(L. 95\%)}}\;\;\text{Mean}\;\;_{\text{(U. 95\%)}} $ \\
\midrule
Conifer&$_{(2.431)}\;\;2.502\;\;_{(2.587)}$&$_{(2.416)}\;\;2.484\;\;_{(2.559)}$&$_{(2.078)}\;\;2.147\;\;_{(2.221)}$\\
 Deciduous&$_{(5.898)}\;\;6.035\;\;_{(6.179)}$&$_{(5.903)}\;\;6.025\;\;_{(6.146)}$&$_{(11.500)}\;\;12.081\;\;_{(12.710)}$\\
 Mixed&$_{(3.689)}\;\;3.852\;\;_{(4.013)}$&$_{(3.687)}\;\;3.825\;\;_{(3.978)}$&$_{(6.792)}\;\;7.307\;\;_{(7.864)}$\\
 Other&$_{(1.555)}\;\;1.617\;\;_{(1.695)}$&$_{(1.533)}\;\;1.597\;\;_{(1.668)}$&$_{(0.315)}\;\;0.340\;\;_{(0.362)}$\\
 \bottomrule
\end{tabular}
}
\caption{Mean canopy height $x_{CH}$ model (\ref{eq:modx}) parameter estimates, with $j$ indexing stratum. Associated process parameter estimates are given in Table~\ref{tab:xfullEstsProcess}.}\label{tab:xfullEsts}
\end{center}
\end{table}

\begin{table}[ht!]
\begin{center}
{\small
\begin{tabular}{lc}
Parameter & $_{\text{(L. 95\%)}}\;\;\text{Mean}\;\;_{\text{(U. 95\%)}}$\\
\midrule
$\nu^2_{CH,0}$&$_{(0.797)}\;\;1.927\;\;_{(6.345)}$\\
 $\nu^2_{CH,TC}$&$_{(0.826)}\;\;1.907\;\;_{(5.836)}$\\
 $\phi_{CH,u}$&$_{(0.560)}\;\;0.585\;\;_{(0.599)}$\\
 $\nu^2_{CH,u}$&$_{(4.666)}\;\;4.823\;\;_{(4.994)}$\\
 \bottomrule
\end{tabular}
}
\caption{Mean canopy height $x_{CH}$ model (\ref{eq:modx}) process parameter estimates. Associated parameter estimates are given in Table~\ref{tab:xfullEsts}.}\label{tab:xfullEstsProcess}
\end{center}
\end{table}

\clearpage
\bibliographystyle{apalike}

\end{document}